\documentclass[]{interact}

\usepackage{epstopdf}
\usepackage[caption=false]{subfig}
\usepackage{comment}
\usepackage{caption}
\usepackage{graphicx}
\usepackage{xcolor}
\usepackage{booktabs}
\usepackage{hyperref}
\usepackage{url}
\usepackage{color}
\usepackage{multirow}
\usepackage{booktabs}
\usepackage[natbibapa,nodoi]{apacite}
\usepackage[longnamesfirst,sort]{natbib}
\bibpunct[, ]{(}{)}{;}{a}{,}{,}
\renewcommand\bibfont{\fontsize{10}{12}\selectfont}

\theoremstyle{plain}

\theoremstyle{definition}

\theoremstyle{remark}

\newcommand{\tip}{{\bf T}}

\newcommand{\alct}{\mathcal{ALC}+\tip_{\bf R}}

\newcommand {\kk} {\mathcal{K}}
\newcommand{\RR}{{\mathcal{R}}}
\newcommand{\TT}{{\mathcal{T}}}
\newcommand{\AAA}{{\mathcal{A}}}
\newcommand{\pfl}{\tip^{\textsf{\tiny CL}}}

\begin{document}


\title{A sensemaking system for grouping and suggesting stories from multiple affective viewpoints in museums}

\author{
\name{Antonio Lieto\textsuperscript{a}\textsuperscript{b}\thanks{Corresponding author Antonio Lieto. Email: antonio.lieto@unito.it}, Manuel Striani\textsuperscript{a}, Cristina Gena\textsuperscript{a}, Enrico Dolza\textsuperscript{c}, Anna Maria Marras\textsuperscript{d}, Gian Luca Pozzato\textsuperscript{a} and Rossana Damiano\textsuperscript{a}}
\affil{\textsuperscript{a}Dipartimento di Informatica, Università di Torino, Italy}
\affil{\textsuperscript{b}ICAR-CNR, Palermo, Italy} 
\affil{\textsuperscript{c}Istituto dei Sordi di Torino (Turin Institute for the Deaf), Turin, Italy}
\affil{\textsuperscript{d}Dipartimento di Studi Storici, Università di Torino, Italy} 
}

\maketitle

\begin{abstract}

This article presents an affective-based sensemaking system for
grouping and suggesting stories created by the users 
about the items of a museum.
By relying on the $\pfl$ commonsense reasoning framework\footnote{$\pfl$ is an acronym for Typicality-based Compositional Logic: the reasoning framework driving the behavior of the the sensemaking system. The framework is described in Section \ref{s:DEGARI}}, 
the system exploits the  spatial structure of the
Plutchik’s `wheel of emotions' to organize the stories according to their extracted emotions.
%
The process of emotion extraction, reasoning and suggestion is triggered by an app, called GAMGame, and integrated with the sensemaking engine. 
Following the framework of Citizen Curation, 
the system allows classifying and suggesting 
stories encompassing cultural items able to evoke not only the very same emotions of already experienced or preferred museum objects, but also novel items sharing different emotional stances and, therefore, able to break the filter bubble effect and open the users’ view towards more inclusive and empathy-based interpretations of cultural content. 
The system has been designed tested, in the context of the H2020EU SPICE project (Social cohesion, Participation, and Inclusion through Cultural Engagement)\footnote{\url{https://spice-h2020.eu/}}, in cooperation the community of the d/Deaf  and on the collection of the Gallery of Modern Art (GAM)\footnote{\url{https://www.gamtorino.it/en}} in  Turin. We 
describe the user-centered design process of the web app and of its components and we report the results concerning the effectiveness of the of the diversity-seeking, affective-driven, recommendations of stories.
\end{abstract}

\begin{keywords}
Story-Based Recommendations; Diversity-seeking emotional recommendations; Commonsense Reasoning; Affective Computing;
\end{keywords}


\section{Introduction}
\label{introduction}




In the last two decades, the awareness of the role of cultural heritage in promoting and enforcing social inclusion has progressively grown, as clearly witnessed by the statements made by institutional actors in various public settings and documents. This trend, started with the FARO Convention \citep{faro2005}\footnote{\url{http://conventions.coe.int/Treaty/EN/Treaties/Html/199.htm}}, culminated with the new museum definition\footnote{\url{https://icom.museum/en/resources/standards-guidelines/museum-definition/}}) released by the Extraordinary General Assembly of the International Council of Museums (ICOM) on 24 August 2022.
Signed in 2005, the FARO convention sees cultural heritage institutions as drivers of reflection and inclusion in society, putting 
``people’s values,  aspirations and  needs first'' and celebrating ``the diversity  and plurality  of their views and values'' \citep{fairclough2014faro}. 
The new definition of museums by ICOM  
puts a strong emphasis on inclusion, diversity and sustainability (``Open to the public, accessible and inclusive, museums foster diversity and sustainability''),  
stressing at the same time the involvement of communities in various activity types, which include education and reflection (``they operate [...] with the participation of communities, offering varied experiences for education, enjoyment, reflection and knowledge sharing"). 

In parallel with the revision of the role of heritage mentioned above, the advent of new technological paradigms, such as mobile technologies, has deeply affected communication and dissemination of cultural heritage\citep{marras2016case}. Technologies can contribute to overcoming the physical barriers interposed between collections and visitors; in addition, today's lower costs of technologies work in favour of their adoption in museums. The COVID-19 pandemic, then, has pushed forward the adoption of technologies to reach an increasing number of audiences with the help of virtual environments.
In the face of these opportunities, changes in audience involvement cannot be limited to the adoption of technologies: on the contrary, technologies call for novel ways to involve and engage visitors and for inclusive design solutions that actually leverage the potential of cultural participation to tackle exclusion \citep{campagnaro2016beauty,giglitto2023digital}. 

%
In this paper, we describe a case study that aims at supporting  reflection on cultural items through the an affective lens, leveraging diversity-seeking recommendations of collections of user-generated artwork interpretations. 
The case study revolves around an app, called GAMGame, for creating and sharing stories about museum collections. The stories created by the museum visitors are classified from an emotional perspective, and this information is used to generate 
diversity-seeking recommendations, aimed at breaking the well known filter bubble effect.
%
%
By relying on the DEGARI 2.0\footnote{DEGARI is an acronym that stands for Dynamic Emotion Generator and ReclassIfier} affective-based reasoner \citep{lieto2022degari}, 
the GAMGame recommends to museum visitors stories encompassing cultural items able to evoke not only the same emotions as the already experienced (or preferred) objects, but also novel items triggering similar or opposite emotions. The ultimate goal of this approach is to open the users' perspective towards a more empathetic and inclusive approach to others' perspectives in cultural heritage.

Developed in cooperation with the Turin Gallery of Modern Art (Galleria d'Arte Moderna, GAM) within the EU H2020 SPICE Project, the GAMGame is targeted at the inclusion of the community of the d/Deaf.
According to World Health Organization (WHO)\footnote{\scriptsize \url{https://www.who.int/news-room/fact-sheets/detail/disability-and-health} }, 
deafness will become a major emergency in the next decades: the WHO report points out that $432$ million adults currently experience a form of disabling hearing loss, and these impairments are expected to involve nearly $2.5$ billion by 2050.
%
%
%
The project puts museums at the center of social inclusion processes by leveraging the novel paradigm of Citizen Curation \citep{bruni2020towards,daga2021integrating}.
Citizen Curation reverses the traditional paradigm of curation, where art interpretation is exclusively entrusted to curators, art historians and critics: citizens are put at the center of the interpretation process, thanks to curation methods which prompt personal responses to art, and promote their sharing across people and communities. 
%
In the SPICE project, Citizen Curation methods -- such as creating personal collections of artworks and attaching personal responses to them -- are supported by a socio-technical infrastructure which allows  visitors to create and share their own interpretations of artworks, and to reflect on other visitors' interpretations.  

The paper is organized as follows. After surveying the main issues concerning museum accessibility and deafness in Section \ref{s:background}, we describe the iterative design and evaluation of the environment for the creation of stories (i.e. the GAMGame) in Section \ref{s:design-GAMGame}. Section \ref{overview} describes the logic and architecture of the sensemaking system representing the inference engine of the GAMGame envoronment, whose evaluation is presented and discussed in Section \ref{s:evaluation}. Conclusion and future work end the paper.

\section{Background and Motivations}\label{s:background}

In this section, we review the relevant notions about museum accessibility and deafness, and we illustrate the paradigm of Citizen Curation that constitutes the overarching conceptual framework of the case study.

\subsection{Accessibility in museums}\label{s:museum-accessibility} \label{ss:accessibility-museums}

The concept of accessibility comes in varying degrees and forms and for some time now it has been associated to the concept of inclusiveness, because, though personal, the visit must be lived without barriers and differences, allowing everyone to access the available content and information. 
According the World Health Organization definition, disability is not only a health problem, but a complex phenomenon, reflecting the interaction between features of a person’s body and features of the society in which he or she lives\footnote{\url{https://www.who.int/health-topics/disability}}. Overcoming the difficulties faced by people with disabilities requires interventions to remove environmental and social barriers. According to \citeauthor{addis2005new}, art, in all its manifestations, is a language and therefore a form of communication, and as such it should be affordable and accessible to all \citep{addis2005new}. Technologies are fundamental tools for involving visitors in museums and  overcoming any kind of barrier: physical, sensory, cognitive, economic, cultural, and social \citep{marras2016case}. To do so, museums are implementing solutions and tools to be increasingly inclusive, thanks to the lower costs of technology. 
However, it is important to highlight that technologies can also be barriers if not implemented in an inclusive way \citep{marras2020bridgingthegap}: to realize inclusive technologies, some institutions have elaborated their own accessibility guidelines (such as the Smithsonian Institute\footnote{\url{https://access.si.edu/}}); most guidelines refer to the ``design for all'' principles\footnote{\url{https://universaldesign.ie/What-is-Universal-Design/The-7-Principles/}}, as exemplified by \citep{timpson2015great}. 

It is understood that any service offered by the museum, such as apps, audio guides, must be inclusive and integrated, as regards the d/Deaf audiences. The key aspect is that d/Deaf people have a strong visual culture ``resulting from the importance of their visual sense to interact with the world'' \citep{martins2016engaging}. The main services offered by the museums are tours in Sign Language and multimedia guides with Sign Languages translations of the museum collection for the d/Deaf audience. Another aspect on which the methods of engagement are developing is that of social communication, following recent research \citep{alnfiai2017social} that has  analyzed $55$ social and communication apps. The authors found that only $6$ had been designed specifically for d/Deaf people. The main function of the apps is allowing users to send video messages and make live video calls and it also provides a variety of technical methods enabling users to communicate, including text, emojis, speech-to-text video, real time communication, privacy, sign language and large size. Also, in this case, it is important to underline that, regardless of the technology used, the appearance of the contents and their comprehensibility is a fundamental aspect. The proliferation of devices such as tablets and smartphones used by many people in their daily lives, pushes towards Bring Your Own Device (BYOD) with access tools increasingly available on various devices, because everyone can make adjustments to suit their own needs and users have increasing expectations that technology should work for them \citep{jankowska2017smartphone}.

\subsection{Deafness in Digital Heritage}  \label{ss:deafness-ch}
\label{deafnessInDigitalHerigate}

Hearing loss is a physiological condition that is generally estimated to be present in ten percent of the general population. However, only proper deafness has a direct impact on language acquisition and, in western countries, affect 1:1000 of the new-born.
Communication modality, skills and attitudes of deaf and hard of hearing people are very diverse, 
due to several individual and context-based factors. Among those, the age of diagnosis, the level of hearing loss, the non-verbal IQ, the use of cochlear implant and/or hearing aids, the age of implant, the presence of additional disabilities, a migrant background, the quality and the quantity of the linguistic input received, the socio-economic status and if they come from a Deaf family or a hearing family. 
Diversity in the communication of the deaf arise from several different factors, however it is possible to outline clear different profiles. 
Some are Sign Language users \citep{spencer2010evidence}, and mostly they
come from Deaf families, where Sign Language is the only and true mother tongue, often from generations.
Sign languages arise almost anywhere there are deaf people \citep{cardona2007lingue}, and their emergence is generally spontaneous when there is a critical mass of deaf people. By definition, deaf people cannot
hear, but have an intact capacity for language, which finds its way of expression in a visual-gestual language, instead of an auditory-phonetic one. They prefer to be called Deaf, with capital letter D, to emphasize that they are a cultural and linguistic minority \citep{lane2011people}, rather than a group of people with a specific hearing impairment. 
They are proud to use Sign Language and ask to recognize it as a minority language, struggling for the right to use it and to access any linguistic contents through it.
On the opposite side there are cochlear implant users, which refuse the idea of being part of linguistic minority and ask for additional and quality speech therapy and advanced technological devices to cope the
disability. They usually can pronounce oral languages (at different level of proficiency) and understand through a mixed of auditory and lipreading skills.
The huge diversity of factors affecting the deaf and hard of hearing people, leads to several mixed profiles, where sign language coexists with oral speech and cochlear implant and the deaf person can use
the two languages at different levels of proficiency and modality, including variations in the use of fingerspelling, lipreading, non-manual components and mouthing \citep{braem2001functions}.

All those factors affect in different ways not only the linguistic outcome, but, crucially, also the level of understanding of any written texts \citep{brueggemann2004literacy}. As a result, oral and written languages represent a barrier in almost every field of the life of deaf persons \citep{goldin2001profoundly}, including museums' cultural offer and any learning environment \citep{spencer2010evidence}. 
International research agrees  that the outcome of this situation of atypical language acquisition is that some grammatical structures of historical-oral languages ``create constant difficulties in the acquisition process by deaf people" \citep{guasti2017language}, ranging from 
phonology and lexicon (which can be poorer and more rigid), to morphology and syntax. 
Accessibility to written and oral forms of verbal languages is therefore one the biggest challenges museums have to face in their pathway to democratize culture. This include using adapted forms of written language,
taking care to use high frequency words and a plain syntax; overusing images and icons to support a quicker and better understanding of textuality; introducing alternative languages and codes, such us Sign
Language, or, with different kind of disabilities, Braille and Alternative Augmentative Communication.

\subsection{Citizen Curation} \label{ss:citizen-curation}

The paradigm of Citizen Curation, developed within the SPICE project, provides a mode of participation in which citizens apply ``curatorial methods to archival materials available in memory institutions in order to develop their own interpretations, share their own perspective and appreciate the perspectives of others'' \citep{bruni2020towards}. 
In the SPICE project, this paradigm is specifically aimed at engaging minority groups that tend to be underrepresented in cultural activities.
%
Citizen Curation can be described as the combination of two processes, namely, Interpretation and Reflection. Although interpretation conceptually precedes reflection, the two processes are not compartmentalized, but, rather, intertwined: reflection builds upon interpretation, but affects subsequent interpretation, forming the continuous process described by \citeauthor{bruni2020towards} \citeyear{bruni2020towards} as the Interpretation-Reflection Loop (IRL). 
The goal of the IRL is twofold: on the one side, to stimulate reflection by exposing citizens to other citizens' interpretations, letting them appraise diversity in responding to artworks; on the other side, to expand the interpretation process as a consequence of the exposition to diversity. 

In the GAMGame, curation takes the form of storytelling, intended as cognitive process oriented to sharing interpretations in a compact, easily processed, universal form \citep{bruner1991narrative,lombardo2012storytelling,bruschi2018digital}. Inspired by the format of social media stories, well known to the target group of the case study (i.e., teenagers and young adults), citizens are stimulated to interact with the collection of the GAM by creating personal stories from the artworks in the collection. Storytelling, here, is not intended simply as the act of selecting and ordering a set of artworks, but implies a deeper, emotional connection with art, in line with the emotional nature of the aesthetic experience: in order to improve the engagement of the participants with the artworks, in fact, they are prompted to express personal reflections and emotions in response to the artworks they include in their stories.
Personal reflections are constrained to a set of themes that have been acknowledged as specifically relevant to the experience of art and the expression of subjectivity by the curators, and widely discussed in the literature \citep{mcadams2018narrative}: memories, thematic connections, and emotions.

In Citizen Curation, emotions are a relevant part of the sensemaking process. Acknowledged as a primary component of the artistic experience for centuries by aesthetics, emotions are an intrinsic component of the the way people experience artistic expression (\citep{van2016implicit,leder2014makes,schindler2017measuring}.
Emotions also provide an universal language through which people convey their experience of art, well beyond words. Despite the differences between  languages, and the influence of cultural factors, emotions own an universal origin: rooted in evolution, they provide the basis for intercultural communication (\citep{ekman1971constants}.
In this sense, emotions can provide a suitable means for connecting people belonging to different groups, intended as culture, age, education, and different sensory characteristics.
Being exposed to the emotions that others feel in response to the artworks, similar o dissimilar, puts the citizens in a situation of perspective taking, i.e., seeing the world from other perspectives (\citep{pedersen2021introducing,djikic2013reading}.
In Citizen Curation,  this approach is intended  to promote empathy, cohesion and inclusion across social groups, in contrast with the current technologies (e.g. like social media or standard  recommender systems) that lead people toward content that fits their own viewpoint, promoting fragmentation and fostering confirmation biases, instead of cohesion, inclusive reflection, and critical thinking.

\section{Towards the GAMGame}\label{s:design-GAMGame}

The design of the GAMGame web app is the result of the cooperation between the University of Turin (AI and HCI experts, museologists) and the GAM museum (museum curators and educators), with the assistance of the Turin Institute for the Deaf (special education experts). Following the paradigms of user-centered design \citep{10.5555/576915}  and of co-design \citep{doi:10.1080/15710880701875068}, the design process has involved the target group, namely deaf people, in all steps, from the collection of the requirements and the development of the prototypes, to the evaluation and redesign phases.

In this section, we describe the design process, which was almost entirely carried out during the COVID-19 pandemic. Although end users and experts were involved along all the process, some tests could only be conducted online due to the lockdowns occurred in 2020 and 2021.  
The timeline was the following: the collection of the requirements and the preparation of the first prototype of the GAMGame occurred between May and October 2020; the prototype was tested online in November 2020 and a new prototype was created from December 2020 to April 2021; the user study on the new prototype was carried out in presence in July 2021; the integration of DEGARI for recommending artworks and stories (Section \ref{overview}), carried out from September 2021 to June 2022, was tested in presence from February 2022 to October 2022 (Section \ref{s:evaluation}).

The outcome of the design is the current version of the GAMGame, which supports the \textit{interpretation} phase of the Interpretation-Reflection Loop envisaged by the paradigm of Citizen Curation described in Section \ref{ss:citizen-curation} by allowing the users to create personal stories from the artworks in the GAM collection. To close the loop, we subsequently integrated into the GAMGame the story recommendation function (Section \ref{overview}), which, together with the unconstrained  navigation of other users' stories, forms the backbone of the \textit{reflection} component of the GAMGame.

\subsection{Requirements  gathering} 

{
The basic tenet for the design of the GAMGame consisted in the choice of storytelling (and visual storytelling in particular) as the method of election for supporting Citizen Curation in the case study. 
%
Storytelling in fact, not only relies on a universal format -- the narrative format -- whose cultural relevance has been acknowledged by psychologists, designers, and media experts \citep{bruner1991narrative,gershon2001storytelling,lombardo2012storytelling}, but is  pervasive in social media (in the form of social media ``stories"), and so both known and appealing for the youngsters. Based on this assumption, the \textit{interpretation} activity in the GAMGame was structured as the  selection of a sequence of artworks from the museum collection to form a personal narrative.

More precise requirements, then, were developed during the co-design of the prototype, carried out in conjunction with the museum curators and the special education experts. During the co-design phase, held online 
through focus groups (via Google Meet)  and achieved through shared project documents (design specifications, interface sketches, literature surveys), two types of requirements were put forth: 
on the one side, museum curators indicated two specific activities, namely \textit{commenting} and \textit{tagging}, as suitable for stimulating the interpretation of the single artworks in the story by younger citizens; on the other side, special education experts provided a list of interface and interaction requirements tailored to the needs of the Deaf. 

The requirements expressed by the curators 
lead to the creation of a console for annotating the artworks (annotation console) included in the story, with 
tools for tagging, commenting and responding emotionally to each artwork.  
Commenting, in particular, was further articulated into three main dimensions (feelings, memories and inspiration), in line with the process of narrative construction of identity described by \citeauthor{mcadams2018narrative}  (\citeyear{mcadams2018narrative}), in order to stimulate the users to provide personal perspectives in the interpretation of artworks.
%

The requirements posed by the special education experts to meet the needs of d/Deaf users concerned both the interface design and the interaction design, and more in general  the overall d/Deaf users experience. In general, it was clearly stated that the use of the app should rely on text as little as possible and be simple and immediate to use. The latter requirement,  motivated by the fact that deafness can be  accompanied by other physical, perceptive and cognitive conditions, led to the choice to design story creation as a rigidly pipelined activity, where backtracking is inhibited, with a minimal number of steps required to obtain a story. 
Specific requirements, then, concerned text, interface and interaction.
\begin{itemize}
    \item Text (in line with the literature surveyed in Section  \ref{deafnessInDigitalHerigate}): 
    \begin{itemize}
        \item The use of text should be limited to the bare minimum, with reference to the atypical acquisition of language described;
        \item Visual codes should be preferred to text whenever possible;
        \item Text, when irreplaceable, should be short and simple (from both lexical and syntactical perspectives). 
\end{itemize}
    \item Interface: 
    \begin{itemize}
        \item The layout should contain the minimum number of elements to help focalization;
        \item The interface should be characterized by a high contrast between salient elements and background;
        \item  Familiar elements and conventions from social and communication media (icons, widgets)  should be reused.
    \end{itemize}    
    \item
    Interaction
     \begin{itemize}
        \item The interaction flow should be maintained simple and predictable;
        \item A low number of steps should be necessary to complete a task;
        \item Interaction should rely on direct manipulation whenever possible, avoiding complex sequences of actions for adding artworks and commenting them.
    \end{itemize}.
\end{itemize}
}

Although most of these requirements were covered, to some extent, by the Web Content Accessibility Guidelines \citep{kirkpatrick_web_2018}, they role in the design process was not secondary, as they orientated the design towards alternatives to text and a general predominance of the visual language. An example of this approach is given by the possibility, for the users, to add emojis to the artworks as a way to express one's own emotional response to the artwork in a visual way, without using words. This function is achieved by simply clicking on the emoji one wishes to add to the painting and dragging from the central position to the desired position. In social media, emojis are similar to a widely used jargon, especially for new generations \citep{DBLP:conf/isdevel/Wolny16}, \citep{barbieri-etal-2018-semeval}, \citep{DBLP:conf/evalita/RonzanoBPPC18}, \citep{DBLP:conf/ranlp/ShoebRM19}, affordable by categories of users who may have difficulties in producing written text on technological devices (such as older people, people with disabilities or children, who generally do not produce long and content-rich texts). 
As reported in \citep{mack2020social}, recent surveys highlight the inclination of Deaf and Hard of Hearing towards visual communication forms in social media, including emojis. The latter in particular, have been described as closer to the type of facial expressiveness which characterizes Sign Languages. 
%
Emojis, also known as smileys or emoticons,  are considered as the best icons to express affective reactions since their metaphor is based directly on the expression of human emotions and they form a good proxy for the intended sentiment \citep{DBLP:journals/behaviourIT/CenaGMV22}, \citep{DBLP:journals/behaviourIT/CenaGGKVW17}.
Concerning the emojis included in the artwork annotation panel (love, curiosity, delight, joy, fear, sadness and disgust), their selection was driven by the museum curators based on their experience with the social media of the institution, and with the preferences of the audience of teenagers.

Finally, a general request from both curators and experts was to create a safe and inclusive space that should not perceived as tailored to the exclusive needs (and so, to the use) of a specific group, but, rather, designed for easy access by all parties \citep{stephanidis2009universal}. This is in line with the notion of inclusion underlying the SPICE project, whose aim is to improve the multiplicity of points of view by obliterating the boundaries between communities.  


\begin{figure}
\centering
\includegraphics[width=1.0\textwidth]{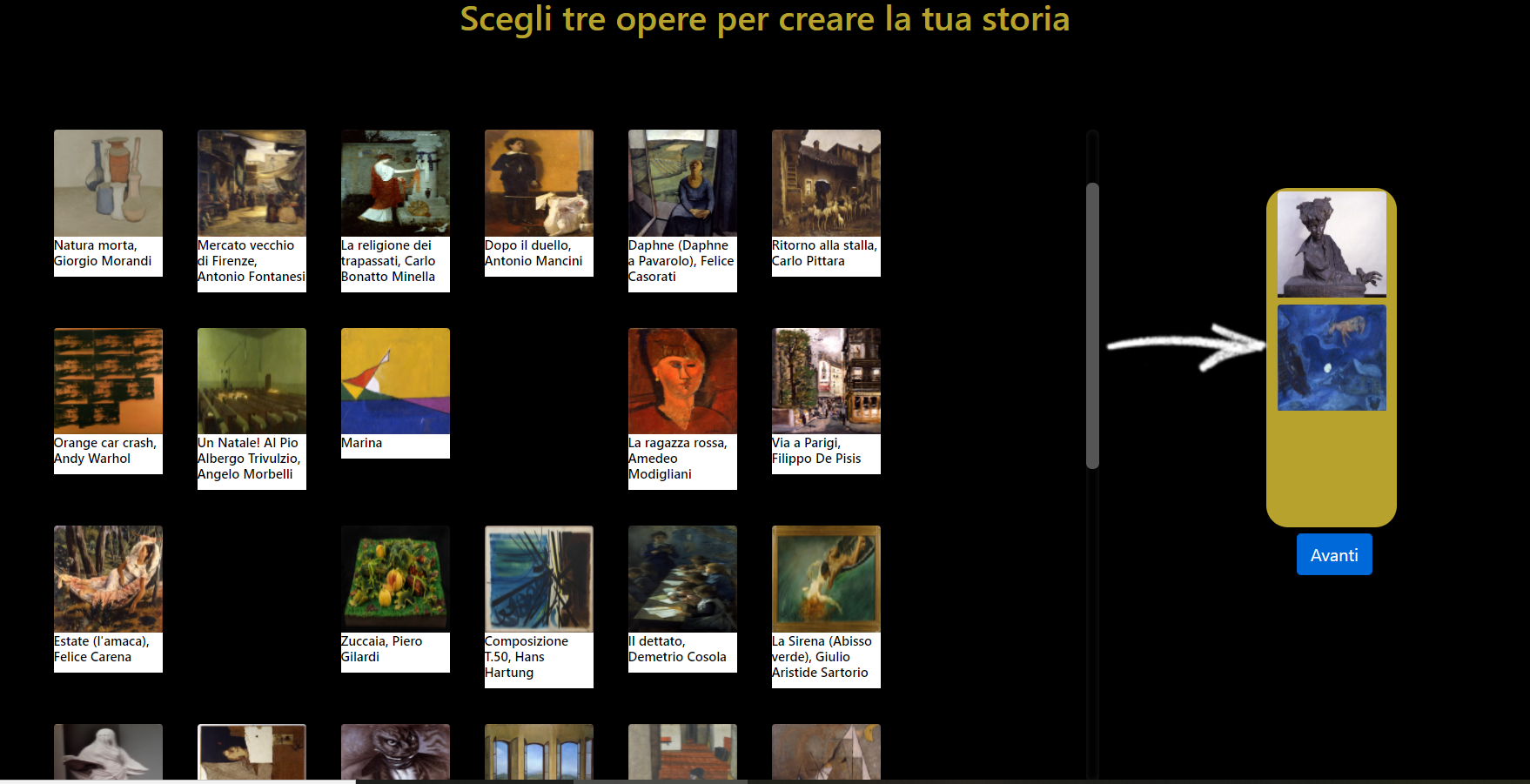}
\caption{A screenshot of the preliminary prototype of the GAMGame web app, representing the first step enabled by the app. Namely: ``Select up to three artworks to create you story".} 
\label{f:old_GAMGame}
\end{figure}

\subsection{Formative evaluation of the prototype}
On November 25-26 2020, while the lockdown in force for the second time, a preliminary prototype of the story creation function (see Figure \ref{f:old_GAMGame}) was tested online with middle and high-school students during a public initiative of the  University of Turin (UNITO) for engaging schools with research (as part of the program of the European Researchers' Night). 
{Given the impossibility of reaching 
the target audience (deaf teenagers and young adults) due to the pandemic, we resorted to the general audience of same age to test the prototype, in a universal design perspective.}
The goal of this test was to confirm the attractiveness of the activity for teenagers, young adults and teachers.

\noindent \textbf{Participants}. A total of $7$ classes ($1$ middle school class, aged 12-14, $6$ secondary schools classe, aged 15-19). 
Since the test was conducted online during the COVID-19 lockdown, it was not possible to identify the participants apart from their school class, due to the characteristics of the meeting platform and the anonymity requirement set by the  event organizers. 

The number of participants ($n$ = $154$) was extracted from the anonymous logs. Participants were distributed by age as follows: $22$ were in the range $11$-$13$, $132$ where in the range $14$-$19$; $15$ were teachers.

\noindent \textbf{Procedure and materials}. Participants took part in the online activity with their teachers in $3$ different sessions, in anonymous form, producing $113$ stories from a catalogue of $43$ artworks selected by the museum curators.

Sessions had the fixed duration of $55$ minutes, since they were part of the school activity program of the classes for the day (both teachers and students were at home and attended school online). 
Sessions were structured as follows: 
after connecting to the Meet platform, 
users received a brief presentation and of the SPICE project by the museum staff (composed of two museum educators) and the university researchers (composed of a museum expert and two scholars in HCI and AI); 
then, the museum educators described the activity, explaining its purpose of creating stories about the artworks.
This introduction was delivered with the help of slides to explain the key ideas of the project; to illustrate the salient phases of the activity (such as artwork selection and annotation), screenshots of the prototype were included, paying attention not to give directions on how to use the interface to accomplish the task. 
After the introduction, users were given the URL of the prototype, which didn't require any installation and could be used online, and were asked to use it individually to create a story of their own.  
The estimated time for the creation of a story was $15$ minutes, but more or less time was allowed according to the need expressed by the students through the chat of the platform.
Teachers supervised the whole session, encouraging the students to participate actively in the activity.
 %

In order to complete a story, users had to select a minimum of $2$  and a maximum of $3$ artworks. 
To advance in story creation, at least one among the available annotation types (tags, comments and emojis) had to be added to each artwork; however, users were free to add as many annotations as they wished. 
$34$ participants were not able to use the web app on their devices, so they conducted the activity by using Google Forms, which had been created in advance in anticipation of technical difficulties.\footnote{$36$ stories were created using Google Forms, but they are not included in the analysis due to the differences with the prototype.}  

\noindent After the activity, we conducted a focus group with the students and the teachers to investigate the users' reception of the activity and to gain insight on its potential for education purposes. 
The focus group was driven by the HCI expert and the AI expert.

During the focus group, participants could contribute to the ongoing discussion by writing their contributions in the video conference chat, or activate their microphone to speak. 
Since sessions could not be recorded for privacy reasons, one of the scholars took note of the questions posed and of the users' contributions.
%
%
The discussion with the users included the following topics: enjoyability of the proposed activity, to investigate its potential to engage the users; 
difficulties encountered in story creation, to eliminate possible obstacles at the interaction and interface level; 
improvement points, to gather suggestions from the users; 
sharing of stories, to determine the users' willingness to share their stories, a crucial step in the Citizen Curation paradigm; 
contexts of use, to explore the onsite and offline use of the app. 
\noindent \textbf{Results}. 
Due to the online mode and the anonymity requirement, 
the quantitative analysis of the results is limited to the collected stories. As mentioned above, $113$ stories were created by the users.   
Stories contained on average  $2,42$ artworks (Standard Deviation = $1,02$); $56$ stories (49.55\%) featured $3$ artworks,  $25$ stories (22,12\%) featured $2$ artworks;  $26$ stories (23\%) contained $1$ artwork, that we attributed to technological issues (client-server communication errors recorded in the prototype logs);  
finally, $6$ stories (5,3\%) featured more than $3$ items. 
 

%
\noindent Some artworks were selected more frequently, although they were presented in random order:
the painting Estate. L'amaca (The hammock)\footnote{\url{https://www.gamtorino.it/it/archivio-catalogo/estate-lamaca/}} was selected $26$ times; Via a Parigi (Street in Paris)\footnote{\url{https://www.gamtorino.it/it/archivio-catalogo/via-a-parigi/}} was selected $21$ times; Le tre finestre (The three windows)\footnote{\url{https://www.gamtorino.it/it/archivio-catalogo/le-tre-finestre-la-pianura-della-torre/}} was selected 15 times. 
Concerning the annotation of artworks, all annotation types were employed by the users (a single annotation of any type was required for each artwork), 
but comments were preferred over emotions, and the latter over tags: 
stories contained in total $136$ comments ($49,82$\%), $73$ emojis ($26,74$\%) and $64$ tags ($23,44$\%).  
Of the comments, $61$ reported feelings ($44,85$ \%), $51$ memories ($37,5$ \%), and $24$ inspirations ($17,65$ \%).
%

\noindent Qualitative data emerged from the feedback provided by the users in the focus group. Here, we report the comments we received to the questions listed above, as they were annotated by the scholars during the focus group. 
Concerning the enjoyability of the experience, there was no negative feedback (here we don't consider the users who were not able to use the online app due to technological problems).
Students reported a very positive appreciation of the experience and some  expressed the wish to repeat it. In particular, they liked the creative aspect of the activity 
(e.g.,``I made a supercute story", $9$ users)  and the possibility of expressing themselves ($16$ users). This is in line with the notion of ``narrative identity'' \cite{mcadams2018narrative}, namely, the creation of an emotional  narrative of self, that underpins the narrative interpretation method in Citizen Curation.
Concerning the difficulties encountered in story creation,  $3$ users declared they were not able to use the drag and drop easily on their devices or that the web app didn't work on MacOs 
(``It doesn't work on my Mac").
%
Concerning the improvement points, $5$ users
asked for a more ample selection of artworks or different artworks 
(``Perhaps changing the artworks to make more stories"), the possibility of adding more than three artworks to the story 
(``Being able to add more than 3 artworks per story", $2$ users), 
and the possibility of applying filters to the artworks and using different colors and fonts when adding tags ($5$ users). Incidentally, the request for typical complements of social media stories (music and photo effects) suggests that the format of social media stories has been acknowledged by these users. 
When asked about the sharing of stories, students declared their  interest in seeing other students' stories, and were ready to share their stories as a condition to see the ones created by the others 
(``I would like to share it", $21$ users), though many of them expressed the wish to do it anonymously 
(``In that case, yes, but I don't want the others to see it with my name", $5$ users). 
Concerning the contexts of use, the feedback was limited to positive answers to the explicit question about the willing to use the app during the museum visit.

 $2$ participants tried both the online prototype and the Google forms 
 and during the discussion expressed a preference for the experience with the web app versus the forms (
 ``I find the form less nice")

\noindent Finally, teachers gave a very positive opinion about the activity and reported that students were able to carry out the activity with minimal assistance (other communication channels, such as Whatsapp class chats, were active during the sessions), with the exception of middle school students, some of which had difficulties in understanding the goal of the activity and we at odds with selecting and engaging with the artworks. In particular, a source of confusion was given by the fact the artwork selection and annotation were separated: first, users selected the artworks to be included in the story, then they were asked to annotate each artwork separately.

%


\subsection{Re-design of the prototype}
Following a design-evaluate-re-design approach, starting from the preliminary evaluation described above, a second prototype was created; while the first one was intended for use on a tablet or desktop computer, suitable for use in parallel with online meetings during lockdowns, the second prototype, developed as a React\footnote{\url{https://reactjs.org/}} web app, is responsive and can be used either on a desktop computer or a smartphone. 
Apart from adapting the interface to the use on mobile phones by doing minor changes (e.g., replacing dragging with clicking for selecting the artworks), the second prototype contained two main changes. 
%
 
%
The first change concerned the selection and commenting order.
Since teachers had reported difficulties by the students in understanding that they had to select in the first place all the artworks they wished to include in the story and then annotate the selected artworks one by one, the creation of stories was broken down into the repetition of selection-and-annotation of each artwork, followed by the assignment of a title to the story before its submission. 
 
\begin{figure}
\centering
\includegraphics[width=1.0\textwidth]{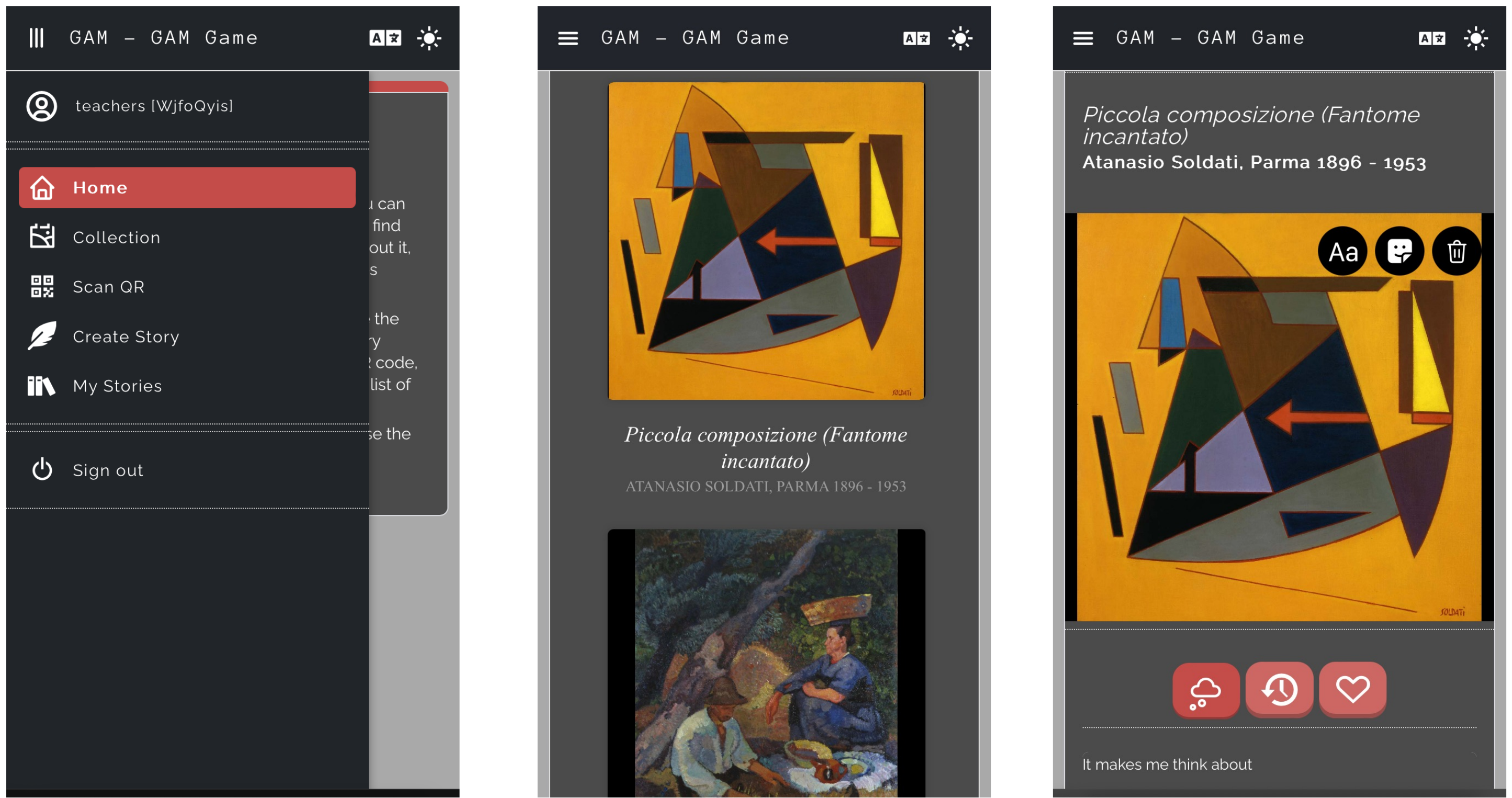}
\caption{GAMGame: Main menu (left), artwork selection (center), artwork annotation (right)
\label{f:new_GAMGame}}
\end{figure}

By selecting the ``Create Story" function from the main menu (Figure \ref{f:new_GAMGame}, left), the user can browse the catalog and select the artworks to include by clicking on them (Figure \ref{f:new_GAMGame}, center). When each artwork is selected, the user is taken to the annotation function 
(Figure \ref{f:new_GAMGame}, right). 
 Since the formative evaluation showed that the users employed all the annotation types (template-based text comments, emojis and tags) included in the first prototype, they were all maintained in the second prototype, but the layout of the annotation interface was redesigned for mobile phones.
The largest part of the screen is occupied by the image of the artwork; the top part contains the annotation console, with commands for adding emojis and tags to the artwork. Once added, emojis and tags can be dragged to the position desired by the user, and discarded if needed. Multiple emojis and tags can be added. By doing so, the artwork becomes an intrinsic part of the creative activity, a  whiteboard on which citizens can express their feelings and ideas about the artwork.
The bottom part of the screen is divided into three tabs (with the selected tab in darker hue), which correspond to the question posed by the museum curators to trigger and drive the interpretative process at a more conceptual level. These questions, suggested by the museum curators and educators, correspond to the personal memories evoked by the artwork, the thematic cues triggered by it, and the feelings it raises. However, in order to comply with the directions provided by the experts for access by deaf users, these questions were i) put in affirmative form ii) expressed in the form of templates to be completed iii) accompanied by evocative icons. The templates, respectively, ``it reminds me of ...'', ``it makes me think of ...'', ``it makes me feel ...'', are intended to act as prompts for users input and can be simply filled by inserting a single word. Since the are no conventional icons for these three types of input, we launched a survey in cooperation with the Turin Institute for the Deaf 
to select the best icons. To do so, for each comment type we proposed two alternatives: one consisted of the most popular icon found by searching the Web with the corresponding keyword (\textit{think}, \textit{feel}, \textit{remember}): the other was proposed by the museum staff. The current icons are the ones emerged from the survey. 


\begin{figure}[ht]
\centering
\includegraphics[width=\textwidth]{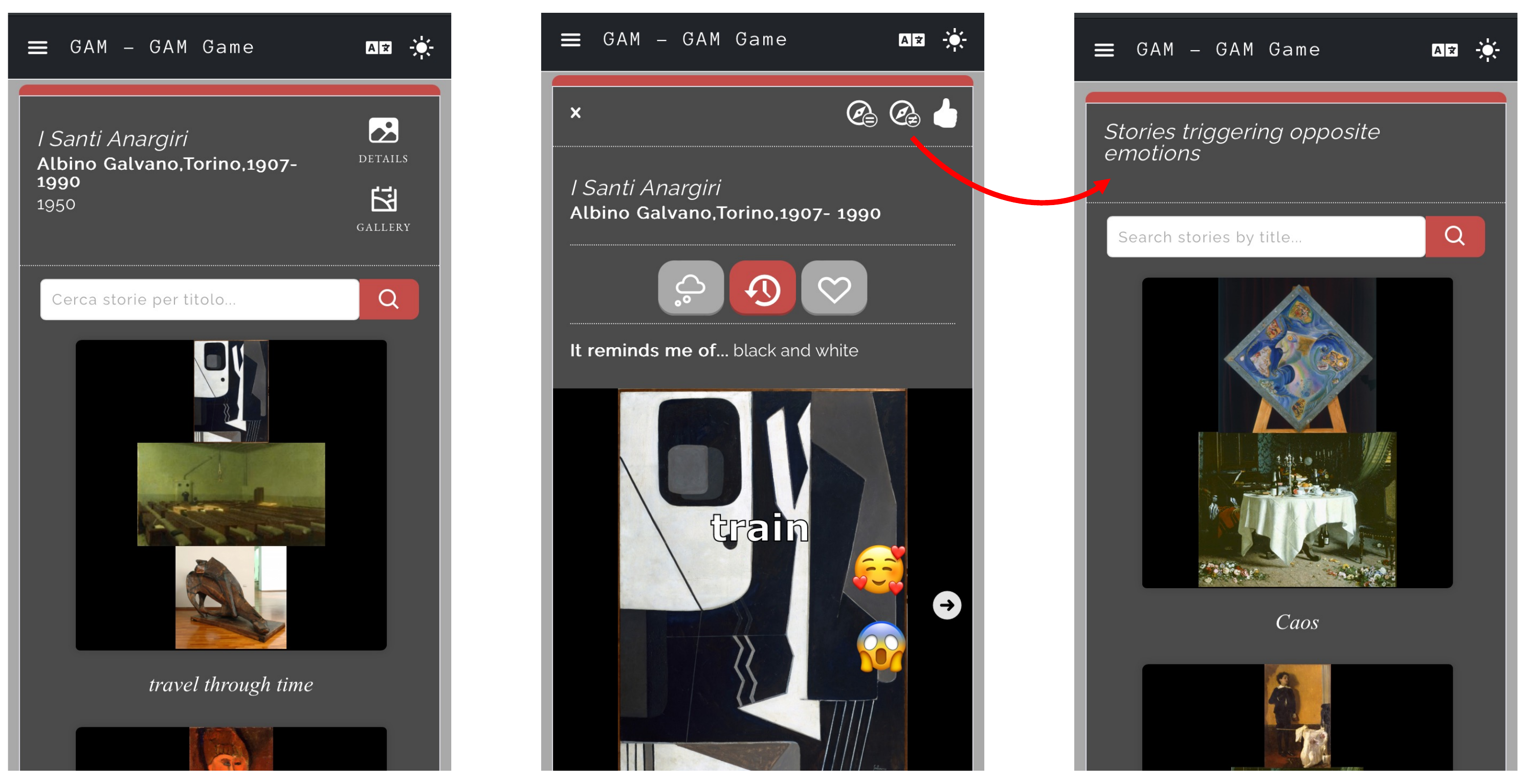}
\caption{The recommendation of stories with similar/opposite emotions. While browsing other users' stories in preview mode (left), the user can  see the annotations of the single artworks in the story (center); for each story, the user can see the stories with similar or opposite emotions (right).} 
\label{f:GamGame-StoryRecommendation}
\end{figure}

More importantly, the function for browsing other users' stories was added. 
in Citizen Curation, seeing other users' stories corresponds to the \textit{Reflection} phase of the Interaction-Reflection Loop described above, aimed at exposing museum visitors to the perspectives of the others. 
Since there no negative feedback was provided by the users in the formative evaluation, we decided to include it in the GAMGame, with the condition that anonymity was preserved as requested by some users.
%
Besides story creation, which corresponds to the \textit{interpretation} phase in the IRL loop described above, the user can explore the stories created by the other users, see her own stories and delete them if she wishes. The exploration of stories in the GAMGame 
is mediated by the museum catalogue: to see the stories stored in the system, the user browses the catalogue and selects an artwork of interested. 
Once the artwork has been chosen, the interfaces shows the link to the stories which contain the artwork. Stories are displayed in preview mode (Figure \ref{f:GamGame-StoryRecommendation}, left); each story can be opened and the artworks in it can be seen, accompanied by the personal annotations added by the user who created it, namely comments, emojis, and tags (Figure \ref{f:GamGame-StoryRecommendation}, center).
Stories can be liked; the stories created by the user are grouped in the myStories section.

\begin{figure}[ht]
\centering
\includegraphics[width=0.8\textwidth]{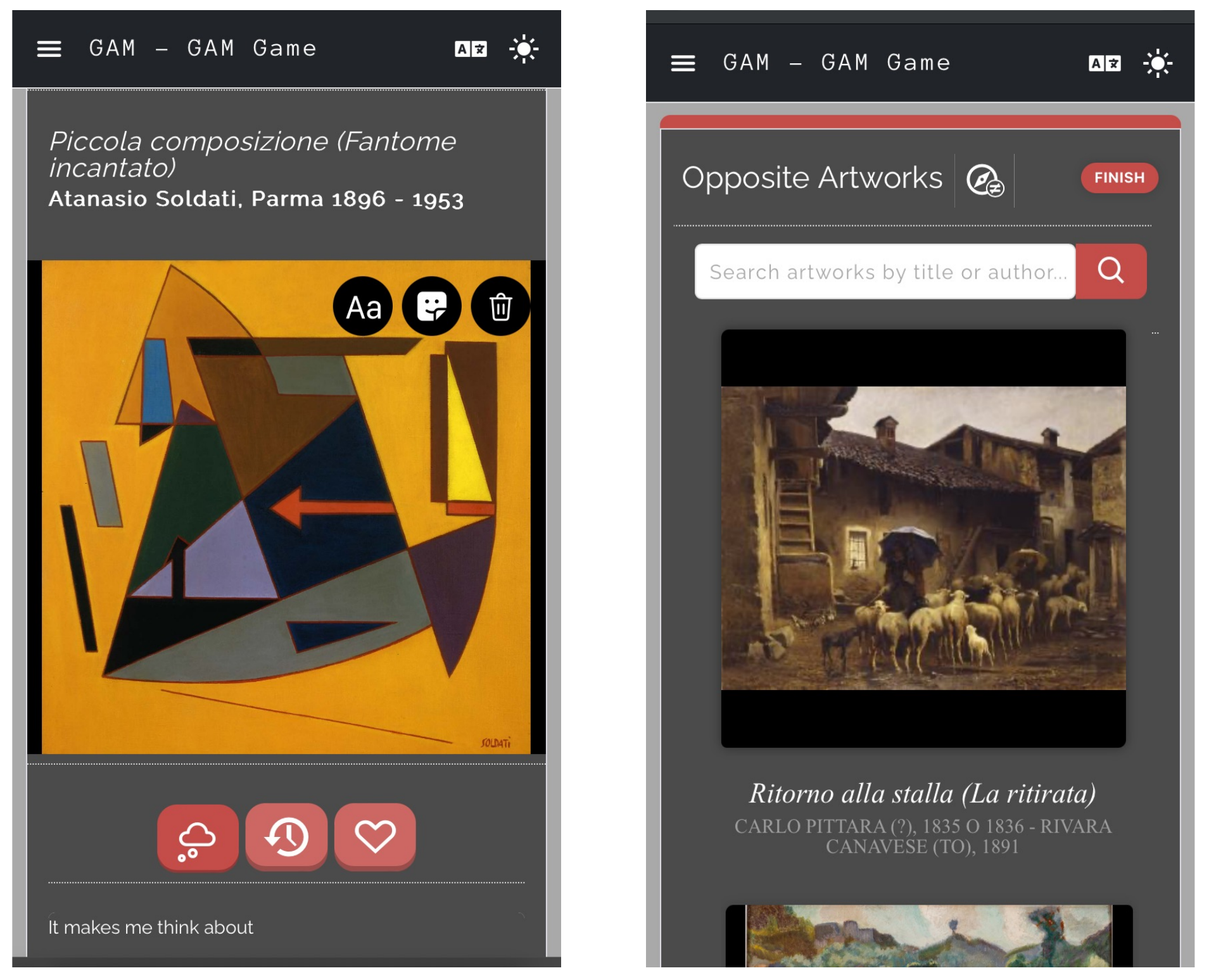}
\caption{The recommendation of artworks in the GAMGame. After selecting the last artwork in the story (left), the user is presented with artworks with similar and opposite emotions (here, opposite, right). Recommended artworks can be added to the story and annotated.} 
\label{f:GAMGame-artwork-recommendation}
\end{figure}

Although the GAMGame web app has been designed with the explicit purpose of realizing the paradigm  of Citizen Curation, it may fall short to enhance diversity in interpretations and, above all, to advance the reflection process beyond the boundaries established by the tendency of the users to search for confirmation of one's choices. In order to improve diversity in both interpretation and reflection, we added emotion-based recommendations to story creation and exploration. To improve diversity, and to leverage the role of emotions in the interpretative process, we relied on the DEGARI 2.0 system \citep{lieto2022degari} to obtain diversity-oriented, affective recommendations from the emotions associate by the users and the curators to the artworks through annotations (users) and artwork descriptions (curators).
%
The generation of emotionally diverse recommendations by DEGARI relies on the Plutchik's model of emotions \citep{plutchik2001nature}, which combines a categorial approach to emotions, with distinct emotion types such as joy, awe or fear, with a dimensional approach that sets emotions into similarity and opposition relations, useful to explore diversity.

In the GAMGame, after creating a story, the user receives a recommendation based on the emotional features associated to the  artworks in the story, as illustrated in Figure \ref{f:GAMGame-artwork-recommendation}). The user can ignore the recommended artworks of both type (with similar and opposite emotions), or can decide to include them in the story (currently, the number of selectable artworks has been limited to $1$ for each recommendation type).
This approach, described by \citep{lieto2022degari} for artwork recommendation, in the present work has been applied to story recommendation, with the goal of orienting to diversity also the reflection phase. 
In the GAMGame, when the user browses the stories in which a given artwork has been included (Figure \ref{f:GamGame-StoryRecommendation}, left), they can select a story and see the single artworks in the story, with their annotations (Figure \ref{f:GamGame-StoryRecommendation}, center). For each story, the systems shows the link to other stories with similar or opposite emotions (Figure \ref{f:GamGame-StoryRecommendation}, right). 
As it will described in Section \ref{s:evaluation}, in our experiments, 
preferred stories with similar and opposite emotions received a better reception over the baseline of stories with the same emotions (which were consequently omitted from the app).

\subsection{Evaluating the usability of the web app}
\label{webappvalidation}
Within the Citizen Curation paradigm, which leverages the citizens'  interpretations of artworks as a way to develop perspective taking, the effectiveness of the socio-technical infrastructure through which the interpretative process is achieved is crucial. 
The annotation of the artworks with personal comments and emotions, in particular, is the core of the interpretation process, and should be accessible to all users in order to ensure the generation of emotionally rich, diverse interpretations.
%
%
For this reason,  in July 2021  we conducted a user study ($n = 12$) to assess  the effectiveness, the user's satisfaction  and 
the perceived ease of use of the story creation function by d/Deaf users, and of the annotation tools in particular. 
Given the issues emerged from the collection of requirements about the use of text by the d/Deaf, we were interested in evaluating the annotation function, which contained textual elements, and we expected the d/Deaf to prefer simpler annotations (tags and emojis) over more complex annotations (text templates). 
At the interface level, we wanted to ascertain that the users would understand the icons that accompany (and in some case replace)  the textual instructions in story creation, some of which had borrowed from social media (e.g., the like icon), while some others had been co-designed with the museum and the  Turin Institute for the Deaf as described above. 
Finally, we wanted to investigate the users' disposition to share their stories, since this is a main pillar of the Citizen Curation paradigm. 

The experiments followed the ethical guidelines released by the SPICE project consortium as part of the Work Package Ethics, ratified by an independent ethics advisor. 
All the user data were anonymously collected and stored according to the project data management plan.\footnote{\url{https://spice-h2020.eu/document/deliverable/D1.2.pdf}
}

\noindent \textbf{Participants}. A convenience sample of $12$ d/Deaf users ($6$ males and $6$ women), selected by the Turin Institute for the Deaf from their staff and students, took part in the experiments. Even though random sampling is the best way of having a representative sample, these strategies require a great deal of time and resources. Therefore, much research in human-computer interaction, in particular for groups of minority, is based on samples obtained through non-random selection \citep{straits2005approaches, young2014approaches}. 
 Following the project ethical guidelines, the selection of  participants was delegated to the Turin Institute for the Deaf, which had the capability to both disseminate the call among its contacts and guarantee a fair selection of the candidates.
The sample intentionally reflected the composition of the community that revolves around the Turin Institute for the Deaf, which includes d/Deaf and non-deaf staff members such as professionals (e.g., Italian Sign Language interpreters, special education experts, media producers), therapists (e.g., speech therapists), teachers (e.g., Sign Language teachers), and care givers, and d/Deaf beneficiaries and trainees. 

%
%
%
As a consequence, the participants belonged to two main groups: 
Group A ($6$ users) consisted of d/Deaf users with other conditions in comorbidity, selected among the trainees. 
As described in Section \ref{deafnessInDigitalHerigate}, in fact, deafness and hearing loss are accompanied by other conditions (such as learning disorders, dyspraxia, etc.) to a higher degree than non-deaf population.
 For this reason, this group is relevant for evaluating the GAMGame in a Design For All perspective.
Group B ($6$ users) consisted of d/Deaf educators and teachers;  this group is relevant for the GAMGame since trainers and educators are expected to plan and drive the use of the web app in educational contexts, within and outside the museum.

%
All participants in group A were aged $19$ to $35$ and had a secondary education level.
In Group B, $4$ participants were in the $19$-$35$ age range, $2$ in the $36$-$60$ age range; $5$ participants out of $6$ had a tertiary education level, $1$ participant had a secondary education level.
Group A included $4$ women an $2$ men; group B included $2$ women and $4$ men. 
Both groups followed the same procedure, with the same apparatus and material, but 
they will be kept distinct in the analysis as a way to improve our understanding of the different parts of the community they represent. 

\noindent \textbf{Procedure}. The user study consisted in the task of creating a story by using the GAMGame app. In order to better control the execution of the activity, story creation was broken down in sub-tasks: artwork selection, artwork annotation and story submission. After receiving brief introduction to the GAMGame objective and features (with no preview of the tool), users were requested to select and annotate $3$ artworks, so the first two steps (artwork selection and annotation) were repeated three times in a row. 

\noindent \textbf{Apparatus}. To avoid disparities in the execution condition determined by the differences between smartphone models, the task was executed by all users on the same desktop computer. Users were assisted by a LIS (Lingua Italiana dei Segni, Italian Sign Language) interpreter, who translated the text to Sign Language if needed, and translated users' comments and questions from LIS to Italian. The screen was recorded during task execution; however, since the experiment was conducted anonymously, the faces of the users were not  included in the shot. During the execution of the task, they were observed by an experimenter taking notes on task completion and details of the interaction with the system.

\begin{figure}
\centering
\includegraphics[width=1.0\textwidth]{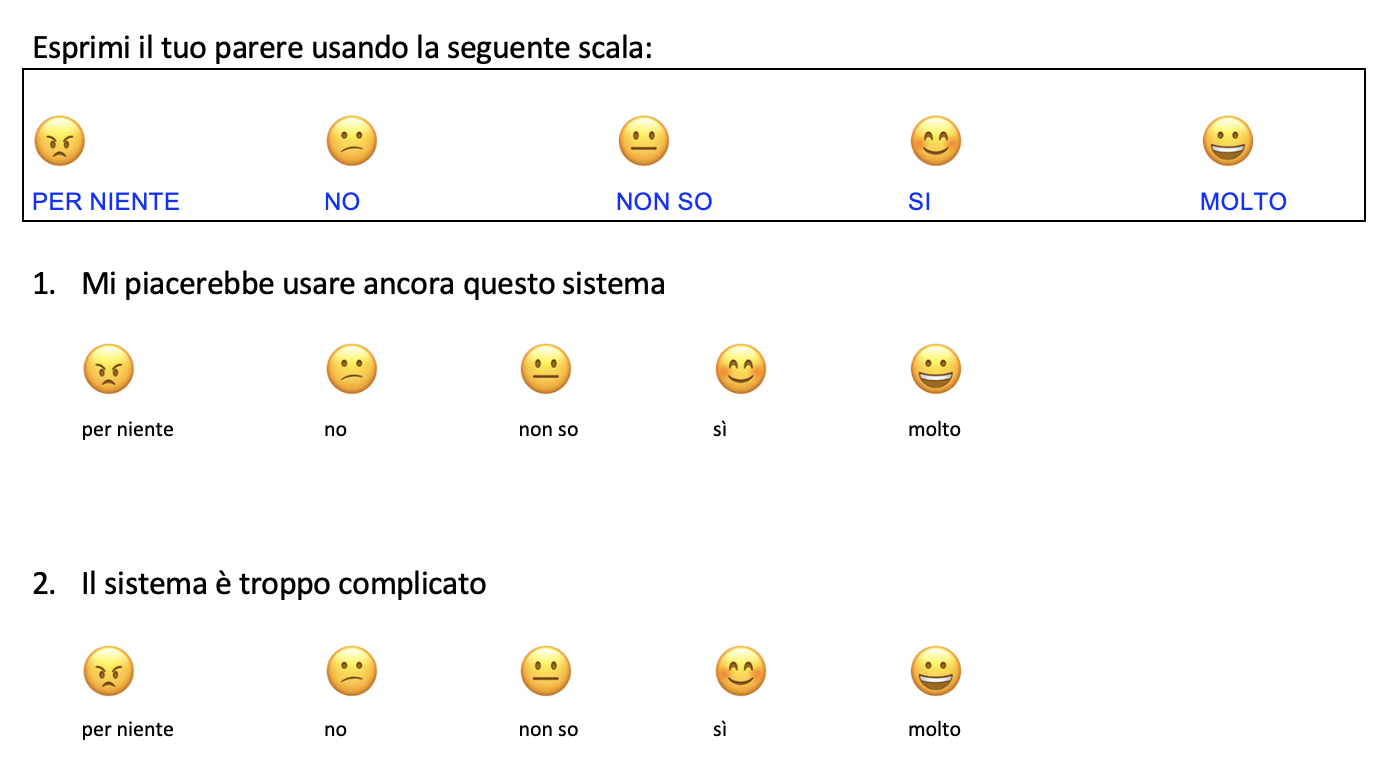}
 \caption{ Use of emojis to collect the user's feedback in the adapted SUS questionnaire (in Italian: ``Express your opinion using the following scale: not at all, no, not sure, yes, absolutely''; ``I'd like to use this system again''). For the sake of space, only the first two questions are shown. During the experiments, instructions and questions were translated by the ISL interpreter.}
\label{f:SUS}
\end{figure}

\noindent \textbf{Material}. In order to  collect information about the socio-demographic characteristics of the users and their use of media, the users, after receiving information about the project and the experiment's objectives, filled out a brief questionnaire on age range, gender, education level, computer literacy and use of social media (Facebook, Instagram, TikTok). 

\noindent After completing the task, users were presented with an adaptation of the System Usabilty Scale (SUS) \citep{brooke1996sus,lewis2018system}: items had been linguistically simplified by putting all questions in affirmative form and replacing difficult words with simpler words of everyday use. The use of Likert scales to collect the answers was replaced by emojis ranging from very sad and sad faces to happy and very happy faces to express agreement, with neutral face as the intermediate item. 
Two extra items were added to the SUS questionnaire, with the goal of collecting feedback on the two aspects mentioned above, namely icons and story sharing: one question was about the familiarity of the user with the icons in the interface (``I found the icons familiar''),  
the other was about the inclination to share one's stories with the other users (``I'd like to share my stories''). These two items, expressed in affirmative form to make them simpler, were added at the end of the SUS questionnaire for convenience, but they were not included in the calculation of the results.


\noindent \textbf{Results and discussion}.
 Concerning the completion of the tasks, we observed a difference between group A (d/Deaf beneficiaries and trainees)
 and group B (d/Deaf teachers and educators)
 (Table \ref{t:tasks}). Users in group A were able to complete all tasks without help; on the contrary, some users in group B needed help to complete the tasks, but they were able to complete the task autonomously after receiving initial help. In detail, concerning the selection of the artworks, $4$ users out of $6$ in group B needed help to select the first artwork, but this number dropped to $1$ for the second artwork; all users were able to complete the selection of the third artwork autonomously. 
 Concerning the annotation of the artworks, the protocol prescribed that the users should add at least one annotation type for each artwork, but they were free to choose which one to use. More than one annotation of the same time could be added to the same artwork (e.g., multiple emojis, comments of different type, or tags).
 As reported in Table \ref{t:tasks}, comment were added $18$ times, tags were added $9$ times, and  emojis were added $23$  times. 
 Proportions slightly change if we consider group A and group B: 
 group A used comments $7$ times (av. number of  comments per task: $2.33$ SD=$0.57$), tags $1$ time, 
 and  emojis $12$ times (average number of  emojis per task: $4$, SD=$2.64$), resulting in general less prolific in annotations ($20$ annotations) than group B ($30$ annotations).
 Group B used comments 11 times (av. number of  comments per task: 3.66 SD=0.57), emojis 11 times (av. number of  emojis per task: 3.66, SD=1.15), and tags 8 times (av. number of tags per task: 2.66, SD=0.57),  with 30 annotations in total. 
 Concerning the differences between the two groups (see Table \ref{t:tasks}), a paired t-test was run on their annotations, showing that the differences between the number of comments (p=0.47) and tags (p=0.007) are significant, while the difference between emojis is not significant  (p = 0.85).

 \noindent Concerning the completion of the tasks, $5$ users out of $6$ in group A requested help to use the comment templates and $5$ out of $6$ requested help for emojis; no user in group B asked for help. 
 To sum up, despite the clear preference for emojis, as expected, the use of text annotation by the d/Deaf contradicted the expectation that tags would be preferred over comments. 
 On the contrary, text comments were preferred to tags by users in Group  A, which represented the most critical group, while no a clear preference emerged for Group B. In our opinion, this preference is due to the fact that the comment templates, curated by museum educators, were more in line with a personal, introspective approach to art, while tags, being unconstrained, were less effective in driving the introspective process; in other words, they offered less guidance. 
 This was confirmed by the observations during the experiment: users -- and users in Group A in particular -- put much effort in making sure that they understood clearly the meaning of the three templates, and were particularly attracted by the template evoking personal memories. Besides confirming the directions provided by the curators, this finding is relevant because it opens to a better access of deaf users to the production of text, which in turn is the input to the tools for extracting sentiment and emotions from users' interpretations in the reflection phase.
 
\begin{table}
\centering
\begin{tabular}{lccc} 
\toprule
         & \textbf{Group A}  & \textbf{Group B}  & \textbf{all}  \\ \hline
         
comments & 7  & 11 & 18   \\ 
tags     & 1  & 8  & 9    \\
emojis   & 12 & 11 & 23   \\
\bottomrule
\end{tabular}
\caption{Annotation of the artworks (comments, tags and emojis) generated by the users in two groups of the user study.}
\label{t:tasks}
\end{table}



\noindent According to the data collected through the questionnaire on media use, all users had a social media account: $10$ users were Facebook users, $8$ were Instagram users; only $1$ user declared using TikTok. All users had at least one social media account. 
Only the users in Group B owned a personal computer; all owned a smartphone.%

\noindent Concerning the feedback collected with the SUS questionnaire, the results show a difference between the two groups.
The comparison between Group A and Group B reveals that Group A (average score $68.33$, slightly above average, standard deviation $14.2$)) evaluated the system more positively than Group B ($64.33$, below average, standard deviation $12,22$), although users in Group A encountered more obstacles in using the system. However, a t-test run on the scores of the two groups showed that the difference is not significant (p=0.23).
We hypothesize that users in Group A, having received more help, eventually underestimated the difficulties of using the system and found it easier to use. On the contrary, users in Group B had, in general, a more critical stance towards their experience and were less satisfied, being at the same time more familiar with technology and possibly endowed with more conceptual tools thanks to a higher education level.

\noindent The rating of the statement on ``Familiarity of icons" ($3,17$ on a 5-point scale,  standard deviation $1,19$) raises some concern on the adequacy of the icons and widgets in the interface and suggests that more work is needed on this aspect, while Item 12 (``Sharing stories", $4,67$ on a 5-point scale, the highest value at all) confirms the positive feedback on sharing one's interpretations received from online users in the previous steps. 

\subsection{Stakeholder Focus Group}



The user study was followed by a focus group that involved the project stakeholders, namely the museum (two museum educators, a curator and the museum's digital officer) and the Turin Institute for the Deaf (a special education expert, a Sign Language interpreter, and two deaf media producers and developers).  According to \cite{abras2004user}, in fact, the successful design of a product must take into account the wide range of stakeholders of the artifact.

In particular, we were interested in the feedback from the museum staff.
The paradigm of Citizen Curation, in fact, does not overshadow the role of curators and educators in museums, who are in charge of  setting the digital environment where citizens' interpretations are created 
and, more importantly, of designing the educational context where these interpretations will be shared and will become the object of reflection. 
In this sense, the experience of museum curators and educators was crucial to assess the potential of our approach and to understand the factors that would hinder the use of the GAMGame.

%
The focus group was conducted by a University team including a HCI expert e two accessibility experts with a background in cultural heritage communication.
The aim of the focus group was to discuss the context of use of the GAMGame and the integration of additional functions for improving the sharing of perspectives between the users (the reflection phase in Citizen Curation terms). 

The museum staff was positive about the integration of the app within their professional practices. 
According to the museum staff, the app should not overlap with the standard museum guide, but at the same tinme should not be relegated to the use at the museum (during or after the visit). According to the museum, the app should be advertised at the museum, with the goal of encouraging people to use it after the visit to share their personal interpretations of the collection. 
By doing so, visitors may return to the visit experience at a later time, and consolidate their engagement with the artworks, and with the museum itself.
In their opinion, this type of use by the public may also improve the relationship of the visitors with the artworks that are usually considered harder to understand, like abstract paintings, because other visitors' interpretations may act as mediators with this type of art better than standard curators' texts. 
The Turin Institute for the Deaf staff observed that, for the app to be used outside the museum, its use should very simple and straightforward: according to them, ideally it should be possible to use to entertain oneself ``while queuing at the supermarket checkout and thinking back to the museum visit" (as one of the curators put it), by relying on a very basic set of instructions.

Concerning the integration of new functions, curators proposed to add recommendations as a way to help the users to orientate themselves in the collections and in the growing set of stories added by the other users (`` how to support both
museums and visitors in exploring and reflecting on the range of
accumulated contributions" in \cite{bruni2020towards}, p.1). From the curators' perspective, emotions were pointed out as particularly appealing for the general public, hypothesizing the creation of an ``emotion-driven visit" and the use of user-contributed stories to create highlights and visit paths. 
In this sense, the visitors' interpretations, thanks to their affective component, may provide the museum with a new tool for increasing their knowledge of how visitors relate with the collections and respond emotionally to them.
Concerning the integration of other media than images, and visual elements in general (such emojis), the Turin Institute for the Deaf staff warned that the use of video, and of Sign Language videos in particular, might contradict the identity of the GAMGame as a universal, neutral and safe space, qualifying it as digitally separated place targeted exclusively at d/Deaf, and consequently hampering its capability to put different communities in touch.

To conclude, the elements acquired through this discussion 
confirmed the decision to introduce the story recommendation function described in the next Section, and gave more foundation to the decision to simplify at most the story creation process, making it repetitive and pipelined as described, and include all the annotations of the artworks in the visualization of stories as a way to give more strength to the reflection process.

\section{Story Recommendation with DEGARI 2.0}
\label{overview}



As described in the previous section, the GAMGame has proved to be a suitable environment for  story creation in an inclusive perspective. Given this background, and the acceptance of affective recommendations of cultural items reported in \citep{lieto2022degari,bolioli2022complementary}, we decided to extend the use of diversity-oriented affective-based recommendations to the recommendation of stories, as way to support perspective-taking and empathy. 

The core component of our affective-based sensemaking system, called DEGARI 2.0, relies on a probabilistic extension of a typicality-based Description Logic called  $\pfl$ (Typicality-based Compositional Logic), introduced in (\citep{jetai}). This framework allows one to describe and reason upon an ontology with commonsense (i.e. \emph{prototypical}) descriptions of concepts, as well as to dynamically generate novel prototypical concepts in a knowledge base as the result of a human-like recombination of the existing ones.

\subsection{Overview of the TCL logic used in DEGARI 2.0}\label{s:DEGARI}

$\pfl$ logic combines three main ingredients. The first one relies on the Description Logic (DL) of typicality $\alct$ introduced in (\citep{AIJ2014})  which allows to describe the \emph{prototype} of a concept.
In this logic, ``typical'' properties can be directly specified  by means of a
``typicality'' operator $\tip$ enriching  the underlying DL, and a TBox can contain inclusions of the form $\tip(C) \sqsubseteq D$ to represent that ``typical $C$s are also $Ds$''.
As a difference with standard DLs, in the logic $\alct$  one can consistently express exceptions and reason about defeasible inheritance as well. 
For instance, a knowledge base can consistently express that ``normally, athletes are fit'', whereas ``sumo wrestlers usually are not fit'' by  $\tip (\mathit{Athlete}) \sqsubseteq \mathit{Fit}$ and $\tip (\mathit{SumoWrestler}) \sqsubseteq  \lnot \mathit{Fit}$, given that $\mathit{SumoWreslter} \sqsubseteq \mathit{Athlete}$. 
The semantics of  $\tip$  is  characterized by the properties of \emph{rational logic}, recognized as the core properties of nonmonotonic reasoning.  As a second ingredient, the logic $\pfl$ exploits a distributed semantics similar to the one of probabilistic DLs known as DISPONTE (\citep{disponteijcai}, allowing to  label inclusions $\tip(C) \sqsubseteq D$ with a real number between 0.5 and 1, representing its degree of belief/probability, assuming that each axiom is independent from each others. As an example, we can formalize that we believe that a typical athlete is fit with degree $0.9$, whereas we believe that, normally, athletes are young, but with degree $0.75$, with the inclusions $0.9 \ :: \ \tip(\mathit{Athlete}) \sqsubseteq \mathit{Fit}$ and $0.75 \ :: \ \tip(\mathit{Athlete}) \sqsubseteq \mathit{Young}$, respectively.  Degrees of belief in typicality inclusions allow to define a probability distribution over \emph{scenarios}: roughly speaking, a scenario is obtained by choosing, for each typicality inclusion, whether it is considered as true or false.

Finally,  $\pfl$ employs a heuristics inspired by cognitive semantics (\citep{hampton1987inheritance} for the identification of a dominance effect between the concepts to be combined: for every combination, we distinguish a  HEAD, representing the stronger element of the combination, and a MODIFIER.  The basic idea is: given a KB and two concepts $C_H$ (HEAD) and $C_M$ (MODIFIER) occurring in it, we consider only {\em some}  scenarios in order to define a revised knowledge base, enriched by typical properties of the combined concept $C \sqsubseteq C_H \sqcap C_M$.

In $\pfl$, given a hybrid KB $\kk=\langle \RR, \TT, \AAA \rangle$  (composed by typical and standard or rigid assertions, i.e. assertion with and without exceptions, as derived from \citep{lieto2017dual}) and given two concepts $C_H$ and $C_M$ occurring in $\kk$, the logic  allows defining a prototype of the compound concept $C$ as the combination of the HEAD $C_H$ and the MODIFIER $C_M$, where  the typical properties of the form $\tip(C) \sqsubseteq D$ (or, equivalently, $\tip(C_H \sqcap C_M) \sqsubseteq D$) to ascribe to the concept $C$ are obtained by considering blocks of scenarios with the same probability, in decreasing order starting from the highest one. Here  all the inconsistent scenarios are discarded, then: (1) we discard those scenarios considered as \emph{trivial}, consistently inheriting  all the properties from the HEAD from the starting concepts to be combined;
   (2) among the remaining ones, we discard those inheriting properties from the MODIFIER in conflict with properties that could be consistently inherited from the HEAD;
   (3) if the set of scenarios of the current block is empty, i.e. all the scenarios have been discarded either because trivial or because preferring the MODIFIER, we repeat the procedure by considering the block of scenarios,  having the immediately lower probability.
 Remaining scenarios are those selected by $\pfl$.
 The ultimate output is a KB in $\pfl$ whose set of typicality properties is enriched by those of the combined concept $C$.
Given a scenario $w$ satisfying the above properties,  the prototype of  $C$ is defined as the set of inclusions $p \ :: \ \tip(C) \sqsubseteq D$, for all $\tip(C) \sqsubseteq D$ that are entailed from $w$ in the logic $\pfl$.

This framework has been applied in a number of applications ranging from computational creativity \citep{lieto2019applying}) to cognitive modelling \citep{lieto2019beyond, chiodino2020goal} and intelligent multimedia, musical and artistic recommendations \citep{chiodino2020knowledge, lieto2021commonsense, lieto2022degari}.

\subsection{Emotion-based Recommandations}\label{s:DEGARI-emotions}

\begin{figure}[]
\centering
\includegraphics[width=1.0\textwidth]{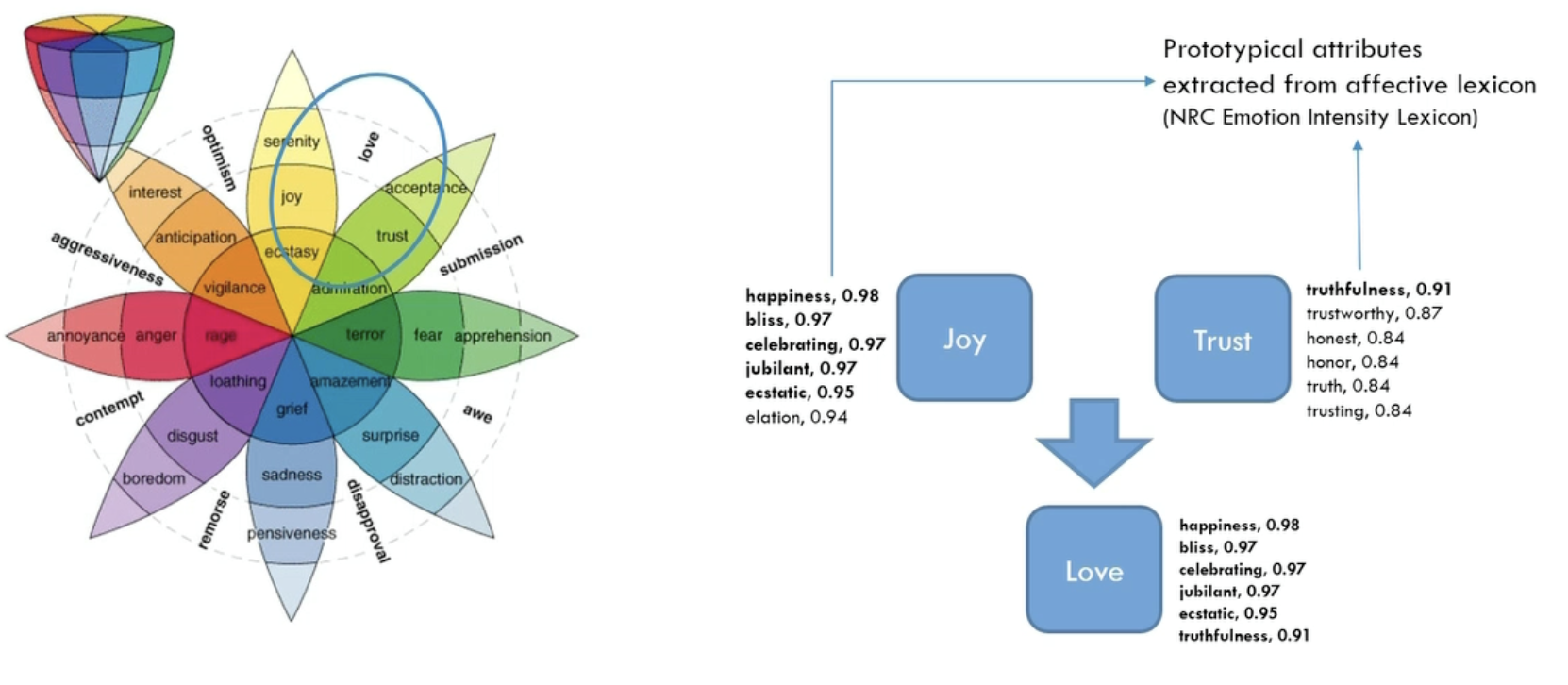}
\caption{Generation of novel Compound Emotions with DEGARI 2.0 by exploiting the Plutchik’s ontology (e.g. Love as composed by Joy and Trust in the picture). The features and the probabilities characterizing each basic emotion are obtained from the NRC affective lexicon. The Plutchik's wheel of emotion in this figure reports only the compound emotions representing the primary dyads, but our system works on the entire spectrum of dyads.
\label{DegariEmotionGenerator}}
\end{figure}

In the context of this work, $\pfl$ is exploited by our system to generate, and classify accordingly, complex emotional concepts (i.e. compound emotions, based on the combination of basic ones) by exploiting an ontological formalization of the circumplex theory of emotions devised by the cognitive psychologist Robert Plutchik \citep{plutchik1980general}, \citep{plutchik2001nature}\footnote{The reasons leading to the choice of this model as grounding element of the DEGARI 2.0 system is twofold: on the one hand, this  it is well-grounded in psychology and general enough to guarantee a wide coverage of emotions, thus giving the possibility of going beyond the emotional classification and recommendations in terms of the standard basic emotions suggested by models like the Ekman's one (widely used in computer vision and sentiment analysis tasks). This affective extension is aligned with the literature on the psychology of art suggesting that the encoding of complex emotions, such as {\em Pride} and {\em Shame}, could give further interesting results in AI emotion-based classification and recommendation systems \citep{Silvia2009LookingPP}. Second, the Plutchik wheel of emotions is perfectly compliant with the generative model underlying the $\pfl$ logic.}. According to this theory, emotions, and their interconnections, can be represented on a spatial structure, a wheel (as reported in the left of the  Figure \ref{DegariEmotionGenerator}), in which the affective distance between different emotional states is a function of their radial distance.
The Plutchik's ontology, formalizing such a theory, encodes emotional categories in a taxonomy, representing: basic or primary emotions; complex (or compound) emotions; opposition between emotions; similarity between emotions. In particular, by following Plutchik's account, complex emotion are considered as resulting from the composition of two basic emotions (where the pair of basic emotions involved in the composition is called a dyad). The compositions occurring between similar emotions (adjacent on the wheel) are called primary dyads. Pairs of less similar emotions are called secondary dyads (if the radial distance between them is 2) or tertiary dyads (if the distance is 3), while opposites cannot be combined\footnote{The ontology is available here: 
\url{https://raw.githubusercontent.com/spice-h2020/SON/main/PlutchikEmotion/ontology.owl}  
and queryable via SPARQL endpoint at: \url{http://130.192.212.225/fuseki/dataset.html?tab=query&ds=/ArsEmotica-core}}. An illustrative example showing the rationale used by DEGARI 2.0 to generate the compound emotions (in this case, the emotion Love as composed by the basic emotions Joy and Trust, according to Plutchik's theory) is reported in Figure \ref{DegariEmotionGenerator}. 

The lexical features associated to each basic emotion (and the corresponding probabilities) comes from the NRC lexicon \citep{mohammad2017word} and, in the context of DEGARI 2.0, represent the prototypical (i.e. commonsense) features characterizing emotional concepts and taken by the system to leverage the $\pfl$ reasoning framework and to generate the prototypical representations of the compound emotions. Once the prototypes of the compound emotions are generated, DEGARI 2.0 is able to reclassify museum items taking the new, derived emotions into account. As a consequence, such a reclassification allows the system to group and recommend museum items based on the novel assigned labels and, as mentioned, a novel prerogative of DEGARI 2.0 consists in the possibility of delivering also diversity-seeking recommendations.


\begin{figure}[]
\centering
\includegraphics[width=0.9\textwidth]{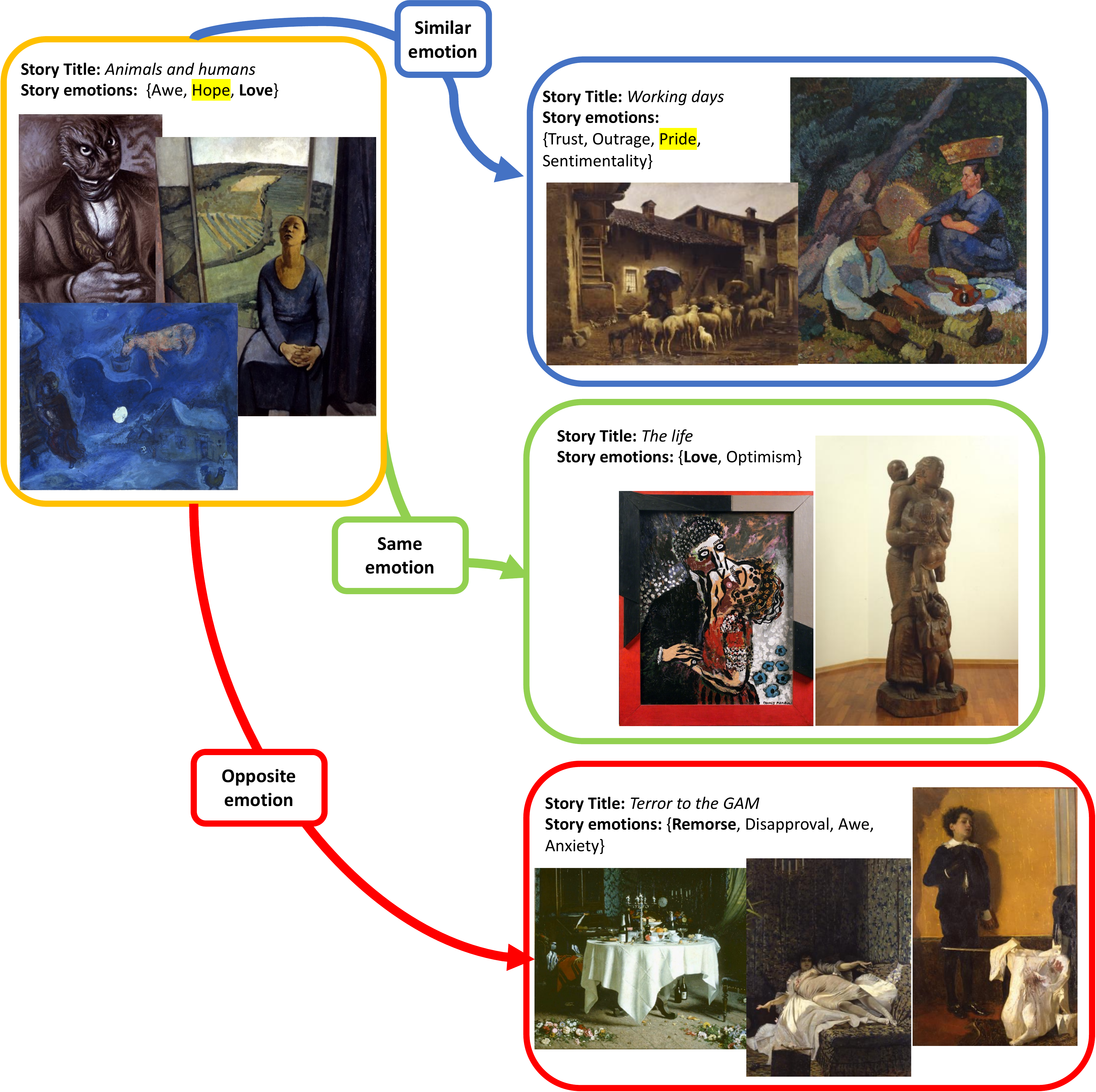}
\caption{
Example of Same, Similar and Opposite emotion for stories recommendations of DEGARI 2.0 from the GAM dataset. This figure shows how the system is able not only to generate new compound emotions but also to group and suggest cultural stories according to their obtained Plutchik's-based affective classification. The entire dyadic structure of the Plutchik's model is  exploited to recommend items and stories (i.e. collections of items) evoking different emotional stances with the aim of providing a more inclusive and affective-based interpretations of cultural content.}
\label{SimilarAndOpposite_GAM}
\end{figure}

The Figure \ref{SimilarAndOpposite_GAM} reports an example of the affective-based story aggregations provided by the system (where a \emph{story} consists of an aggregation of selected museum items). Here, the emotion ``\textsl{Hope}'' (with background-yellow) detected in the story entitled ``Animals and humans'' triggers the similar emotion ``\textsl{Pride}'' (with background-yellow) in the story ``Working days''. The emotion ``\textsl{Love}'' (in the first story, highlighted in bold) triggers the emotion ``\textsl{Love}'' (highlighted in bold) in recommended story (with same emotion) entitled ``The life''. Finally, the emotion ``\textsl{Love}'' (in the first story, highlighted in bold) triggers the emotion ``\textsl{Remorse}'' in the recommended story (with opposite emotion) entitled ``Terror to the GAM''. The connection between the different types of emotions, and therefore between each story associated, is provided by the Plutchik's ontology exploited by DEGARI 2.0.
Overall, the system tries to categorize and link the stories with respect to any of the original emotional categories found in the forming items.

As anticipated, $\pfl$ is adopted in DEGARI 2.0 to automatically build the prototypical representations of the compound emotions according to the Plutchik's theory and the information about the emotional concepts and their corresponding features to combine via $\pfl$ are extracted from the NRC Emotion Intensity Lexicon (\citep{mohammad2017word}\footnote{Such lexicon provides a list of English words, each with real-values representing intensity scores for the eight basic emotions of Plutchik's theory. The intensity scores were obtained via crowd-sourcing, using best-worst scaling annotation scheme.}. This lexicon associates words to emotional concepts in descending order of emotional intensity and, for our purposes, we considered the most intensively associated terms for each basic emotion as typical features of such emotion. In this way, the prototypes of the basic emotions were formed, and the $\pfl$ reasoning framework is used to generate the compound emotions.  
Such prototypes of basic emotions are formalized by means of a $\pfl$ knowledge base, whose TBox contains both \emph{rigid} inclusions of the form $$\mathit{BasicEmotion} \sqsubseteq \mathit{Concept},$$ in order to express essential desiderata but also constraints, as an example $\mathit{Joy} \sqsubseteq \mathit{PositiveEmotion}$ as well as \emph{prototypical} properties of the form $$p \ :: \ \tip(\mathit{BasicEmotion}) \sqsubseteq \mathit{TypicalConcept},$$  representing typical concepts of a given emotion, where $p$ is a real number in the range $(0.5,1]$, expressing the frequency of such a concept in items belonging to that emotion: for instance, $0.72 \ :: \ \tip(\mathit{Surprise}) \sqsubseteq \mathit{Delight}$ is used to express that the typical feature of being surprised contains/refers to the emotional concept {\em Delight} with a frequency/probability/degree of belief of the $72\%$.

Once the association of lexical features to the emotional concepts in the Plutchik's ontology is obtained and the compound emotions are generated via the logic $\pfl$, the system is able to reclassify the cultural items in the novel formed emotional categories. Intuitively, an item belongs to the new generated emotion if its metadata (name, description, title) contain all the rigid properties as well as at least the $30\%$ of the typical properties of such a derived emotion. The 30\% threshold was empirically determined: i.e., it is the percentage that provides the better trade-off between over-categorization and missed categorizations \citep{chiodino2020knowledge}.        


\subsection{DEGARI 2.0 Software Modules and Architecture}
\label{degariarchitecture}

\begin{figure}
\centering
\includegraphics[width=1.1\textwidth]{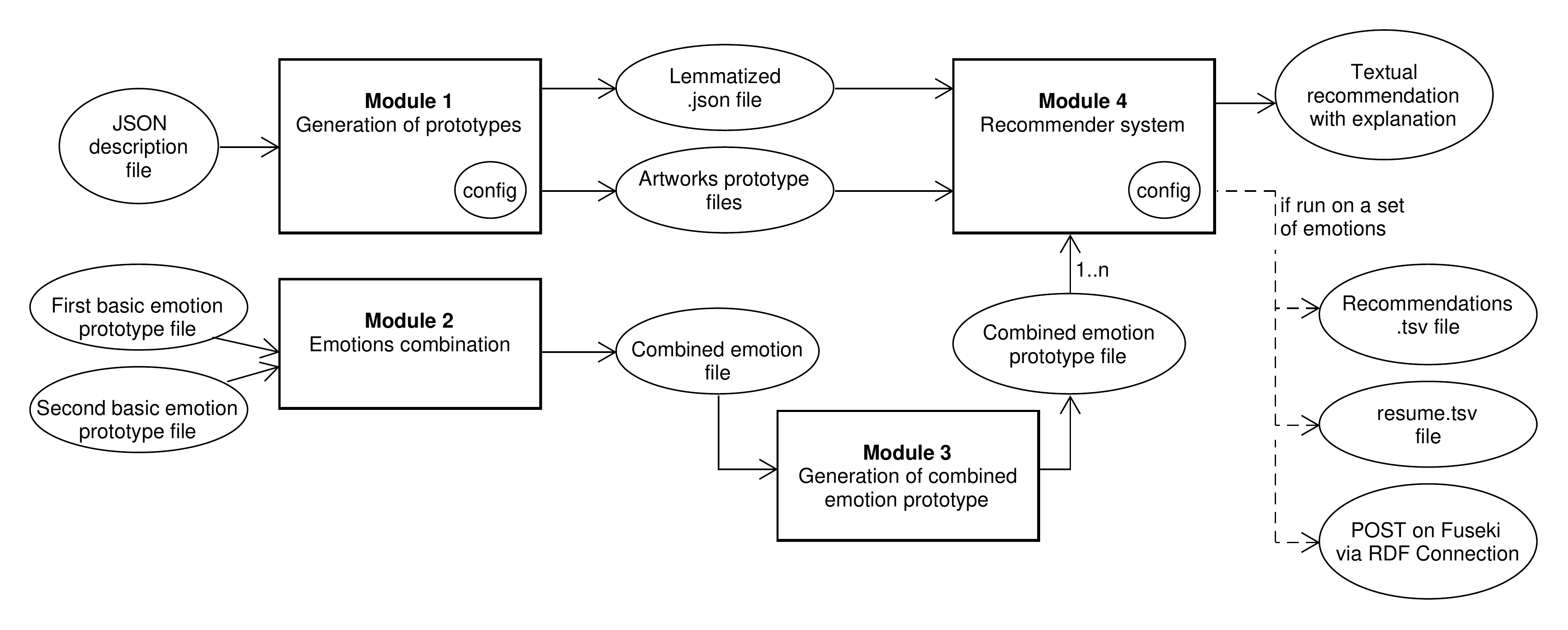}
\caption{The software architecture of DEGARI. Module 1 represents the entry point of the system. It accepts JSON files containing a textual description of the cultural items (coming from user comments or from the museum catalogues) and performs an automatic information extraction step generating a lemmatized version of the JSON descriptions and a frequentist-based extraction of the typical terms associated to the cultural item. Modules 2 and 3 are devoted respectively i) to the acquisition of the basic Emotions to combine (Module 2) and ii) to the generation of the compound Emotions (Module 3). Module 4 is the one classifying, grouping and recommending the cultural item according to the novel generated emotions.}
\label{DegariArchitecture}
\end{figure}

Overall, the system is composed by four software modules,   as depicted in Figure \ref{DegariArchitecture}. The modules adopting  $\pfl$ and involved in the processes of (basic) emotion formation and (compound) emotion generation correspond to the Modules 2 (Emotion combination) and 3 (Generation of combined emotion prototypes) of the architecture in the Figure. Module 1 (Generation of prototypes), on the other hand, represents the entry point of the system and manages the metadata associated to each museum item. Finally, Module 4 (Recommender system), is the one devoted to group, reclassify and recommend the cultural items according to the novel emotional labels created by DEGARI 2.0. 
In particular, the reclassification step requires matching the output of Module 1. Namely: matching the extracted metadata of each museum item (or the user-generated texts associated with it), with the ones characterizing the compound emotions generated in Modules 2 and 3. 

\begin{figure}
\centering
\includegraphics[width=1.13\textwidth]{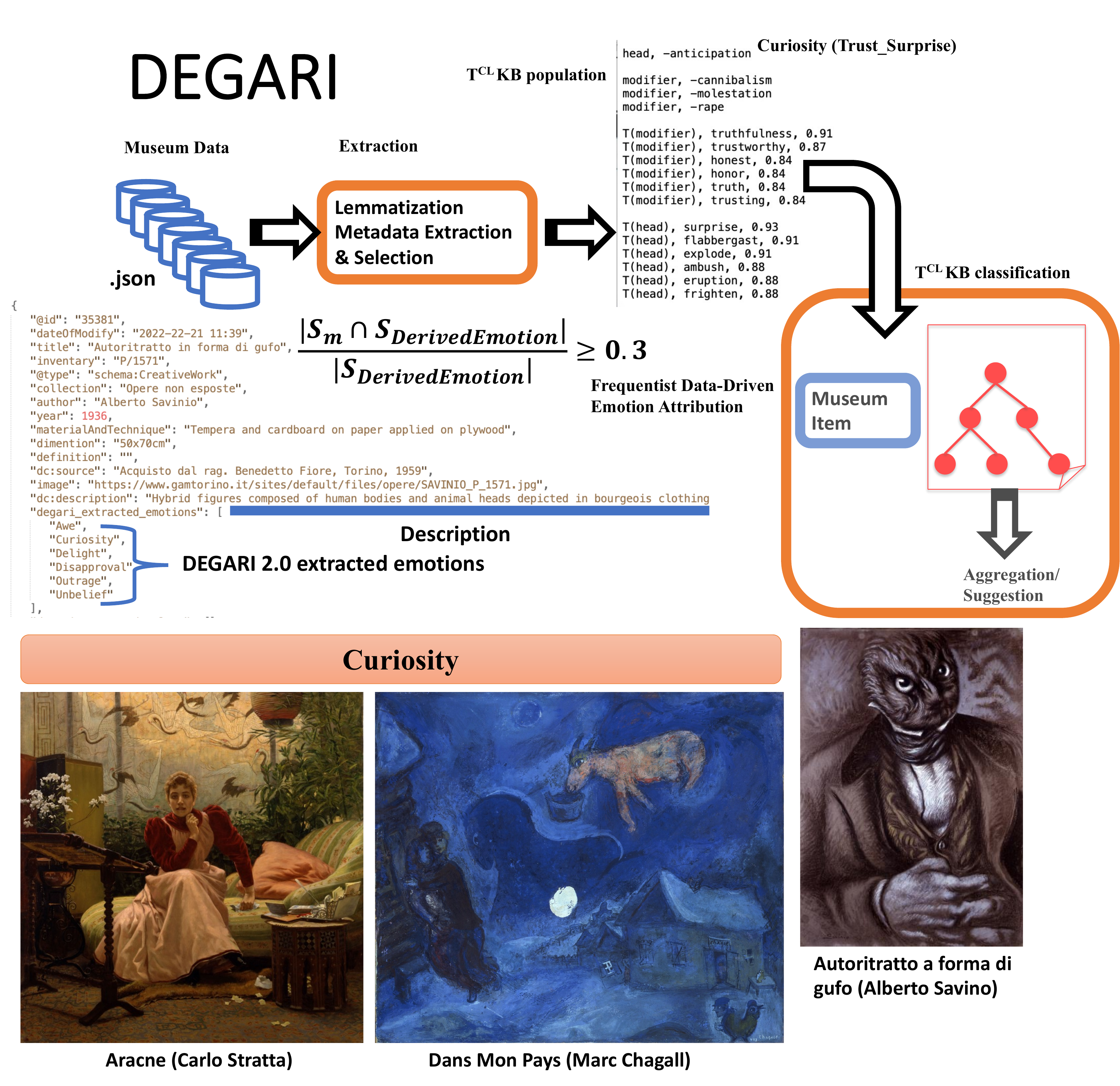}
\caption{A pictorial example of the categorization pipeline used by DEGARI 2.0 for emotion attribution and content aggregation/suggestion based on the artwork ``Autoritratto a forma di Gufo'' (Self-portrait in the shape of an owl) from the GAM. The item is associated with a textual description coming, in this case, from the museum collection (user-generated contents are also handled via the same format). The figure shows the mechanisms, described in the text, for the extraction, classification and grouping of the museum items according to the $\pfl$ generated complex emotional categories (e.g. Curiosity in the example). These depicted steps (focusing on single items  classification and grouping) represent the prerequisite for the recommendations of stories (composed by a collection of items).} 
\label{DegariJsonDescription}
\end{figure}

In the current version of the system, Module 1 accepts JSON files containing a textual description of the cultural items (e.g. coming from user generated comments or from the museum catalogues) and performs an information extraction step generating a lemmatized version of the JSON descriptions of the cultural item and a frequentist-based extraction of the typical terms associated to each cultural item in its textual description (the assumption is that the most frequently used terms to describe an item are also the ones that are more typically associated to it). The frequencies are computed as the proportion of each term with respect to the set of all terms characterizing the item. These two tasks (lemmatization and frequency attribution) are performed by using standard libraries like Natural Language Toolkit\footnote{\url{https://www.nltk.org/}} and TreeTagger\footnote{\url{https://www.cis.uni-muenchen.de/~schmid/tools/TreeTagger/}}.
Once this pre-processing step is done, the final representation of the cultural items is compared with the representations of the typical compound emotions obtained in Module 3. This comparison, and the corresponding classification, is done in Module 4 that implements, we recall, the following categorization heuristics: if a cultural item contains all the rigid properties and at least the 30\% of the typical properties of the compound emotion under consideration, then the item is classified as belonging to it.
After the categorization has taken place, DEGARI is eventually able to classify and group together the items evoking the same emotions (e.g., Curiosity in the Figure \ref{DegariJsonDescription}, this JSON snippet is the element entering the Module 1 of the system and triggering its entire processing until the recommendation step (in this case based on the classical ``same-emotion'' suggestion).) or, as shown in the examples from the Figure \ref{SimilarAndOpposite_GAM}, aggregation of items (i.e. stories) having opposite or similar emotions. 


The current version of the system is available as a web service that that can be invoked via standard HTTP requests and whose reasoning output is made automatically available to a queryable SPARQL \footnote{\url{https://www.w3.org/TR/rdf-sparql-query/}} endpoint.
As we will show in the next section, this advancement allowed us to call the DEGARI 2.0 reasoning services and to integrate its output within a web app (called GAMGame) built to collect user data on cultural items, during a museum visit.

\section{Evaluation of recommendation strategies} 
\label{s:evaluation}
For the experiments, we relied on a an selection of $56$ artworks from the GAM catalogue (which contains $586$  items overall) made by the curators of the Museum for the inclusion in the app. The rationale behind their selection was to present the audience with a variety of subjects, styles, techniques and historical periods so as to overcome curatorial biases in the setting.
The artwork metadata in the GAM catalogue contain the descriptions of  each artwork including title, author, date, characters, actions, objects. 
These elements were encoded as JSON-LD files, so that the resulting description of each item was compliant with the input of DEGARI 2.0 system.

\subsection{Evaluation approach}


To the best of our knowledge, there is no available evaluation standard to test a system about its diversity-seeking affective recommendation (not only referred to the deaf community). As a matter of fact, indeed, standard recommendation systems are evaluated on their ability to confirm one own's points of view. On the other hand, the purpose of a system like DEGARI 2.0 is exactly the opposite: i.e. to break the filter bubble effect by adopting  an inclusive approach aiming at extending (not confirming) the typology of experienced cultural items through the exploitation of an affective lens trying to include, in the user's perspectives and potential experience, also cultural items that do not directly fit their usual, expressed, preferences. 
Given this state of affairs, we tested the capability of DEGARI 2.0 to suggest alternative, diversity-seeking, emotion-based aggregations of museum items  selected by the users. Such aggregates are called \emph{stories}, since they provide a narrative context to the chosen selection\footnote{The analysis of the recommendations based on stories represents the major difference with a previous work \citep{lieto2022degari} that was, on the other hand, focused only on singe-items diversity-seeking recommendations}.

Our evaluation has been carried out in two phases. First: during the researchers' night event (promoted by the University of Turin) and involving both the community of deaf people engaged via 
the Turin Institute for the Deaf as well as non-deaf people. Then, after two weeks, a follow-up was conducted with a subset of the deaf people only (the participants of this second phase had already participated to the researcher's night experiment and were recontacted by the Institute). 

Overall, the experimentation was done on 83 users who created a total of 87 stories using the GAMGame web app. In particular, a group of 34 non-deaf users (8,2\% female,  47,1\% male, 14,7\% not specified 
, who generated 34 stories and a second group of 49 deaf users (55\% female, 40\% male, 5\% not specified ), who generated 52 stories, were involved in the two phases. All participants gave their written
informed consent before participating in the experimental procedure, which was approved by the ethical committee of the University of Turin, in accordance with
the Declaration of Helsinki (World Medical Association,
1991). Participants were all naıve to the experimental procedure and to the overall aims of the study.

In the context of our work, the evaluation of recommendation strategies was done separately from the evaluation of interaction, since it has been inspired by the so-called layered evaluation approaches of recommender/adaptive systems \citep{Brusilovsky:01,Karagiannidis:00, paramythis2005decomposition}. According to these approaches,  during the evaluation of adaptive and recommender systems,   instead of evaluating the system as a whole, at least two layer should be distinguished: the interaction layer, wherein the effectiveness of the changes made at
the interface are evaluated,  and content layer, wherein the accuracy of the system inferences are evaluated. These two layers should be distinguished instead of evaluating the systems as a whole because if the recommender solutions do not improve the interaction, for instance, it is not evident whether one or both the above layers have been unsuccessful. The effectiveness of such an approach has been demonstrated by several experimental results, see \cite{gena2005methods}  for details. The following sections explore the experiments involving the content layer.
 
\subsection{First Experiment}








In the first experiment, both groups of deaf and non-deaf users were asked to make an evaluation on the recommended stories characterized by same/similar/opposite emotions extracted from DEGARI 2.0 on the basis of the story they created.

\begin{figure}[h]
    \centering   
    \includegraphics[width=\textwidth]{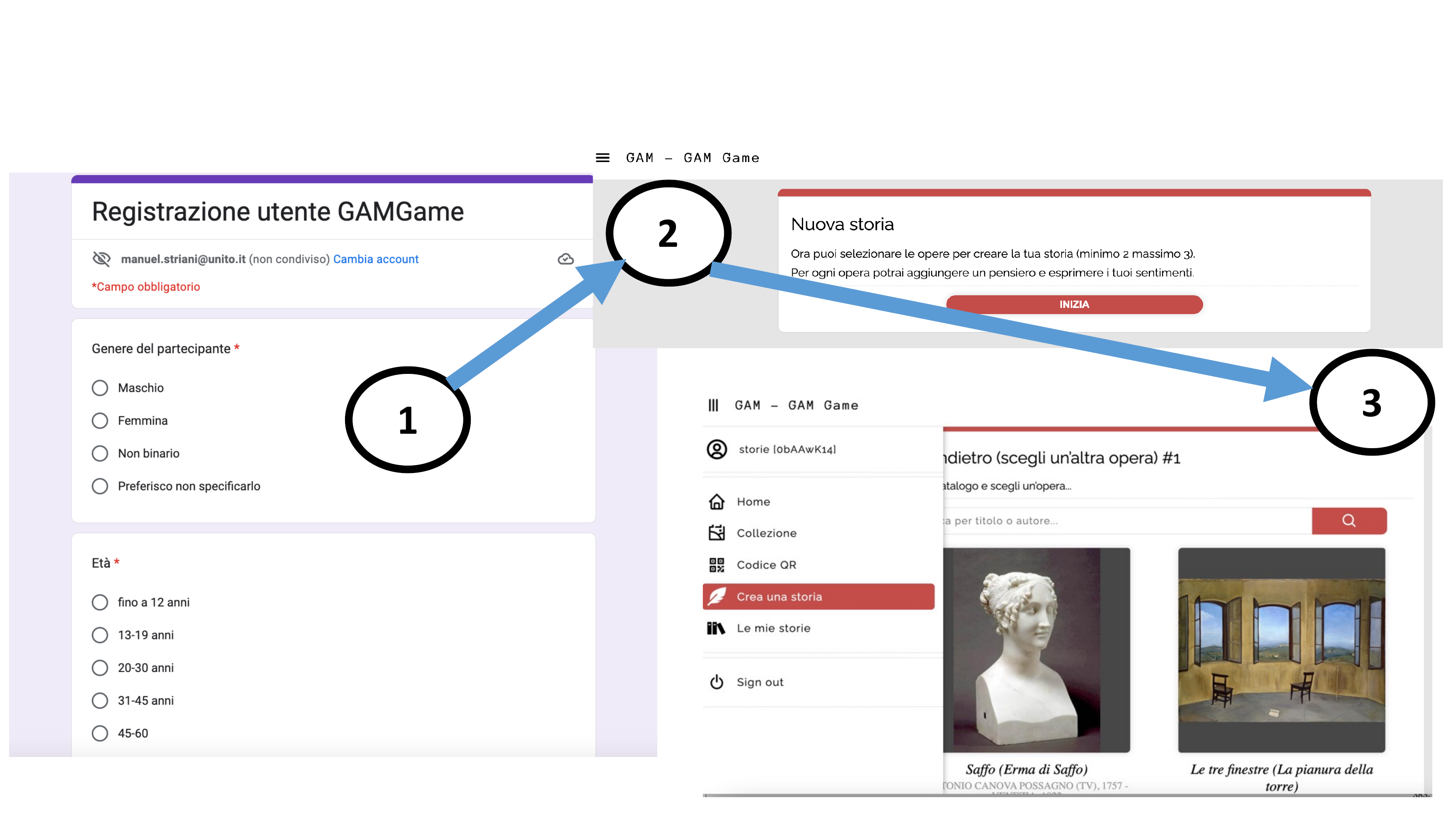}
    \caption{Google-form for user registration to the GAMGame web app during the UNITO researchers' night. On the left, the Google Form with the anonymous data (here, gender and age group); on the right, the link to the GAMGame and the home page (here, for desktop screen) of the GAMGame.}
    \label{UserRegistration}
\end{figure}

\noindent \textbf{Hypothesis}. We have hypothesized that the recommendations receiving the better ratings would be the ones suggesting stories created by other users with exactly the same  emotions as the stories created by the user (e.g. If the extracted emotion associated to a story created by a given user is Love, then we hypothesize that the recommendation of other stories eliciting Love would receive better rating than the one eliciting similar or opposite emotions).

\noindent\textbf{Experimental Design}. Single factor within subjects design. The independent variable
was the recommendation manipulated according to three levels: the received recommendations based on the ``same-emotion'' - based recommendation, the  ``similar-emotions'' based recommendation, and the ``opposite emotions'  - based recommendation.

\noindent \textbf{Participants}. This first evaluation involved  34 non-deaf users (8,2\% female,  47,1\% male, 14,7\% not specified) and 40 deaf users (45\% female, 50\% male, 5\% not specified). The two groups generated 35 and 34 stories respectively.
%
%
Participants were selected by involving the community of d/Deaf people engaged via 
(the Turin Institute for Deaf), via an availability sampling strategy. The reported evaluation focuses on the acceptability of the received inclusive-based affective recommendations.

\noindent \textbf{Procedure}. The experiment aimed at measuring the satisfaction of the potential users of the GAMGame web app when exposed to the suggestions of the novel categories suggested by DEGARI 2.0. It consisted in a user study\footnote{This is one of the most commonly used methodology for the evaluation of recommender systems based on controlled small groups analysis, see \citep{shani2011evaluating}.} where participants, after having provided their own stories in the GAMGame app, were exposed to a number of affective-based recommendations based on their original selection.  
The participants registered via Google Forms; the registration ended with the generation of an anonymous id for the test session. For each (anonymous) user, in the registration phase, we asked to provide the following information: Gender, age, relationship with art, with museum, and if not. Once completed the registration phase (Figure \ref{UserRegistration}), users were redirected to the GAMGame web app where they could   start the creation of their stories. After completing the story, each participant was recommended with three types of stories: stories featuring artworks with the same emotions, stories featuring artworks with similar emotions, and stories featuring artworks with opposite emotions

\noindent\textbf{Apparatus}. The participants used their own devices to fill out the anonymous registration procedure and carry out the story creation task, in compliance with the  practice known as Bring Your Own Device, well established also in the field of museum applications\citep{ballagas2004byod}.

\begin{figure}[h]
\centering
\includegraphics[width=\textwidth]{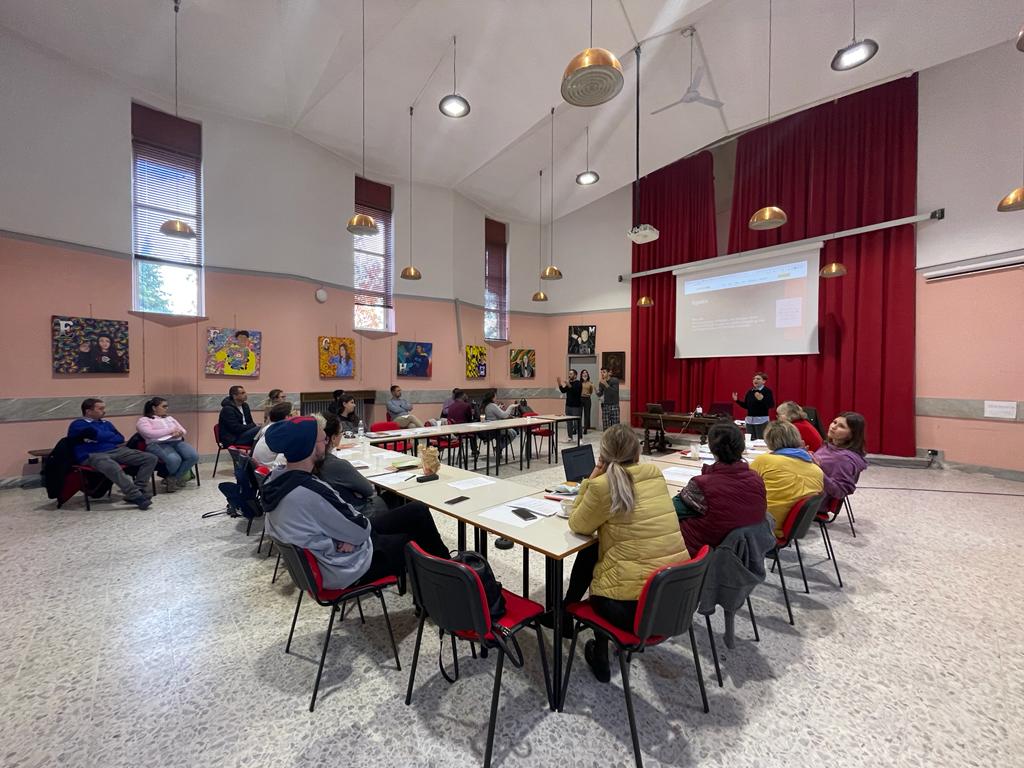}
\caption{A professional translator from Istituto dei Sordi explains to the participants how to use the app for creating their stories.}
\label{Deaf2}
\end{figure}

\noindent\textbf{Material}. 
At the end of the interaction, participants were asked to compile an online questionnaire about the received suggestions. Here, they had to rate, on a 10-point scale (from 1 to 10,  being 1 the lowest and 10 the highest rating), the received recommendations based on the ``same-emotion'' ``similar-emotions'' and ``opposite emotions'' categories.
The instructions to the community of d/Deaf participants were provided and supervised by a professional translator of the Institute for the Deaf (the Figure \ref{Deaf2} shows a frame of this experiment) who translated questions, feedback, and comments from Italian to LIS (Lingua Italiana dei Segni, Italian Sign Language) and vice versa.



\subsubsection{Results and Discussion for the Inclusive Recommendations, First Experiment}
\label{ResultsAndDiscussion}




Overall, the two groups rated 53 recommendations. This rating results are shown in Table \ref{TableRaitingDeafNonDeafUsers}(a) for non-deaf participants who gave their assessment on a set of 24 recommended stories and Table \ref{TableRaitingDeafNonDeafUsers}(b) for d/Deaf-participants participants who gave their assessment on a set of 29 recommended stories.

After the interaction with the GamGame web app, and after having being exposed to the affective-based, diversity seeking recommendation by the DEGARI engine, the d/Deaf and non-deaf participants were asked to make an evaluation on the recommended stories characterized by same/similar/opposite emotions extracted from DEGARI on the basis of the story they created. The overall obtained results about the ratings are shown in Table \ref{TableRaitingDeafNonDeafUsers} (a) for non-deaf participants and Table \ref{TableRaitingDeafNonDeafUsers}(b) for deaf participants.

\begin{table}
\caption{(a) shows the rating for of non-deaf  participants and (b) d/Deaf participants users the DEGARI 2.0 stories recommendations}
\label{TableRaitingDeafNonDeafUsers}
\centering
\begin{footnotesize}
\begin{tabular}{lrrr} 
\toprule
\multicolumn{3}{l}{\textbf{Rating on recommended stories by the non-deaf users}}                                                &                                                          \\
             & \multicolumn{1}{l}{}                             & \multicolumn{1}{l}{}                                   & \multicolumn{1}{l}{}                                     \\
             & \multicolumn{1}{l}{\textbf{Rating same emotion}} & \multicolumn{1}{l}{\textbf{Rating on similar emotion}} & \multicolumn{1}{l}{\textbf{Rating on opposite emotion}}  \\
Mean         & 6,0208                                           & 7,4583                                                 & 7,3125                                                   \\
Dev.Standard & 2,3335                                           & 1,8233                                                 & 1,8871                                                   \\
Median       & 7                                                & 8                                                      & 8                                                        \\
             & \multicolumn{1}{l}{}                             & \multicolumn{1}{c}{(a)}                                & \multicolumn{1}{l}{}                                     \\ 
\hline
\multicolumn{3}{l}{\textbf{Rating on recommended stories by d/Deaf}}                                            &                                                          \\
             & \multicolumn{1}{l}{}                             & \multicolumn{1}{l}{}                                   & \multicolumn{1}{l}{}                                     \\
Mean         & 5,7414                                           & 6,6638                                                 & 5,4483                                                   \\
Dev.Standard & 2,4445                                           & 2,4406                                                 & 2,6937                                                   \\
Median       & 6                                                & 7                                                      & 5                                                        \\
             & \multicolumn{1}{l}{}                             & \multicolumn{1}{c}{(b)}                                & \multicolumn{1}{l}{}                                     \\
\bottomrule

\end{tabular}
\end{footnotesize}
\end{table}

Table \ref{TableRaitingDeafNonDeafUsers} shows the mean, median and standard deviation values for each emotion recommendation group (same, similar and opposite emotions).  It is possible to note that, contrary to our original hypothesis and expectations, both groups (deaf/non-deaf participants) manifest a major preference for the suggestions of stories that generate similar emotions: deaf users (Table \ref{TableRaitingDeafNonDeafUsers}(b)) have an average on the same emotions of  6.66 while the group of non-deaf users (Table \ref{TableRaitingDeafNonDeafUsers}(a)) reaches an average of 7.46 

Concerning the recommendation of stories with opposite emotions, on the contrary, we observe that, while non-deaf users give  with higher ratings to stories with opposite emotions than same emotions, deaf users give higher ratings to stories with the same emotions than opposite emotions. Interestingly enough, the reported data about the story-based affective recommendations for the deaf users confirms the findings reported in a previous experimentation involving only single-items recommendations, see \citep{lieto2022degari}, page 11, Figure b.


\begin{table}[t]
\centering
\begin{tabular}{cccccc} 
\toprule
\multicolumn{6}{c}{{\textbf{Rating on same emotion}}}                                                                                                                                                                                                                                                                                  \\ 
\bottomrule
\multicolumn{1}{l}{}                                                & \multicolumn{1}{l}{\textcolor[rgb]{0.2,0.2,0.2}{{Rating}}} & \multicolumn{1}{l}{$U-value$} & \multicolumn{1}{l}{\textbf{$Z-score$}} & \multicolumn{1}{l}{\textbf{$p-value$}} & {significant/not}                                                                \\
\begin{tabular}[c]{@{}c@{}}Sample 1\\(deaf-users)\end{tabular}      & {22}                                  & \multirow{2}{*}{277.5}                   & \multirow{2}{*}{1.46032}                 & \multirow{2}{*}{0.1443}                  & \multirow{2}{*}{\begin{tabular}[c]{@{}c@{}}Not~significant\\at $p < 0.05$\end{tabular}}  \\
\begin{tabular}[c]{@{}c@{}}Sample 2 \\(non-deaf users)\end{tabular} & 31                                                                &                                          &                                          &                                          &                                                                                         \\
\multicolumn{1}{l}{}                                                & \multicolumn{1}{l}{}                                              & \multicolumn{1}{l}{}                     & \multicolumn{1}{l}{}                     & \multicolumn{1}{l}{}                     & \multicolumn{1}{l}{}                                                                    \\ 
\bottomrule
\multicolumn{6}{c}{\textbf{\textbf{Rating on similar emotion}}}                                                                                                                                                                                                                                                                                                    \\ 
\bottomrule
\begin{tabular}[c]{@{}c@{}}Sample 1\\(deaf-users)\end{tabular}      & 22                                                                & \multirow{2}{*}{342.5}                   & \multirow{2}{*}{0.60789}                 & \multirow{2}{*}{0.54186}                 & \multirow{2}{*}{\begin{tabular}[c]{@{}c@{}}Not~significant\\at $p < 0.05$\end{tabular}}  \\
\begin{tabular}[c]{@{}c@{}}Sample 2\\(non-deaf users)\end{tabular}  & 31                                                                &                                          &                                          &                                          &                                                                                         \\
\multicolumn{1}{l}{}                                                & \multicolumn{1}{l}{}                                              & \multicolumn{1}{l}{}                     & \multicolumn{1}{l}{}                     & \multicolumn{1}{l}{}                     & \multicolumn{1}{l}{}                                                                    \\ 
\bottomrule
\multicolumn{6}{c}{\textbf{\textbf{\textbf{\textbf{Rating on opposite emotion}}}}}                                                                                                                                                                                                                                                                                 \\ 
\bottomrule
\begin{tabular}[c]{@{}c@{}}Sample 1\\(deaf-users)\end{tabular}      & 22                                                                & \multirow{2}{*}{234}                     & \multirow{2}{*}{2.06853}                 & \multirow{2}{*}{0.03846}                 & \multirow{2}{*}{\begin{tabular}[c]{@{}c@{}}Significant\\at $p < 0.05$\end{tabular}}      \\
\begin{tabular}[c]{@{}c@{}}Sample 2\\(non-deaf users)\end{tabular}  & 31                                                                &                                          &                                          &                                          &                                                                                         \\
\bottomrule
\end{tabular}

\caption{Statistical comparison of the rating groups for Sample 1 (rating for deaf users) and Sample 2 (rating for non-deaf users) on same/similar/opposite emotions by using the Mann-Whitney statistics.}
\label{StatisticsSample1AndSample2}
\end{table}

For the statistical comparison (shown in Table \ref{StatisticsSample1AndSample2}) of the rating groups of Sample 1 (rating for deaf users) and Sample 2 (rating for non-deaf users), we used the Mann-Whitney statistic. 
Specifically, for Sample 1 Rating on same emotion for deaf users) and Sample 2 (rating on same emotion for non-deaf users) we collected 22 ratings and 31 ratings respectively, thus obtaining a $U-value$ = 277.5, a $Z-score$ = 1.46032 and a $p-value$ = 0.1443. The results are not statistically significant at $p<$0.05.
For Sample 1 (evaluation of deaf users on similar emotions) and Sample 2 (evaluation of non-deaf users), we thus obtained a $U-value$ = 342.5, a $Z-score$ = 0.60789 and a $p-value$ = 0.54186. The results are not statistically significant at $p<$0.05.
Finally, for the opposite emotion of Sample 1 (deaf users' evaluation of the opposite emotion) and Sample 2 (non-deaf users' evaluation), we thus obtained a $U-value$ = 234, a $Z-score$ = 2.06853 and a $p-value$ = 0.03846. The results are statistically significant at $p<$0.05, confirming that the two groups actually differ in relation with the recommendation of stories with opposite emotions.

The recommendations that received a better rating were the ones suggesting stories linked to the original one through the property ``similar emotion''. The recommendations of stories evoking opposite emotions, with respect to the original created story in the game (for deaf users),  were the ones that received the worst ratings. 
In particular, this latter datum suggests that there are mechanisms of cognitive resistance that prevent a full acceptance of suggestions going in a different direction from  one's own preferences. This datum also suggests that a first guideline that can be extracted for the improvement of diversity-seeking affective recommenders concerns the opportunity to adopt presentation devices for the \textbf{mitigation of cognitive resistance effects}. Although the search for mitigation measures that wrap diversity into some meaning frame is an open research area, the effectiveness of narrative formats \citep{wolff2012storyspace, damiano2016exploring} and of ethically-driven digital nudging techniques  \citep{augello2021storytelling,gena2019personalization} is worth exploring.

A more immediate strategy that could be adopted in our system is also represented by the progressive recommendation of items evoking emotions that are gradually more distant from the starting one (where the distance can still rely on the radial structure of the Plutchik's wheel encoded in the ontology).

\subsection{Second Experiment}

The second experiment involved only a subset of the deaf people that had participated to the previous experiment contacted the Turin Institute for Deaf two weeks after the Researcher's Night.  Here we wanted to evaluate if a repeated exposition to diversity-seeking affective recommendations (i.e. the ones based only on similar and opposite emotions) would have lead (if any) to any variation in the assigned ratings.  


\noindent \textbf{Hypothesis}.  We have hypothesized that the devised framework (formed by the GAMGame interface plus the devised sensemaking engine) presents an overall layout that, after a repeated exposition to the tool,  favours/increases the willingness to the exploration of stories addressing multiple and diverse viewpoints within the museum exhibition. More specifically: we hypothesize that a repeated interaction with the system would lead to an increased preferences (if compared to the first interaction) for the recommendations of stories based on similar and opposite emotions.  

\noindent\textbf{Experimental Design}. The same methods as the first experiment were used for the second one. Single factor within subjects design. The independent variable
was the recommendation manipulated according to two levels: the received recommendations based on  the  ``similar-emotions'' based recommendation, and the ``opposite emotions'  - based recommendation.\color{black}

\noindent \textbf{Participants}. This second evaluation involved only deaf users: $9$ users (5 females, 4 males),   engaged via the Turin Institute for Deaf accepted to participate to the second experiment and generated 18 stories.

\noindent \textbf{Procedure}. The users were asked to create stories by avoiding to select the same items already chosen in the first experiment (with the aim of not being re-exposed to the same recommendations already received). Once logged, in fact, users could see their own previous stories in the My stories section of the GAMGame. The analysis conducted in this experiment can, therefore, be considered as a sort of repeated session where we recorded and compared the ratings assigned to the recommendations with respect to the previously obtained ones. 


\noindent\textbf{Apparatus}. As for the first experiment, the participants used they own devices to fill out the anonymous registration procedure and carry out the story creation task.

\noindent\textbf{Material}. 
As for the first experiment, at the end of the interaction, participants were asked to compile an online questionnaire about the received suggestions. Here, they had to rate, on a 10-point scale (from 1 to 10, being 1 the lowest and 10 the highest rating), the received recommendations based on the ``similar-emotions'' and ``opposite emotions'' categories.
The instructions to the community of d/Deaf participants were provided and supervised by a professional translator of the Institute for the Deaf who translated questions, feedback, and comments from Italian to LIS (Lingua Italiana dei Segni, Italian Sign Language) and vice versa.





\subsubsection{Results and Discussions for Inclusive Recommendations, Second Experiment}

In total, 18 brand-new stories were created and received diversity-seeking recommendations based on stories provided by other users.

The main figure emerging from this experiment is reported in Table \ref{TableDynamicalNudging_ForDeaf}, which compares the average rating assigned by the deaf users for both recommendation types. It emerges, for both the types of recommendations, an augmented preference assigned to the received recommendations, thus suggesting that the repeated exposition to novel points of view can improve  the willingness to widen one's own view. At time $t_1$, in fact, we observe that average rating assigned to the recommended stories with similar emotions grows (from 6.664 at $t_0$ to 7.278 at $t_1$), overcoming the ratings assigned to stories with same emotions -- as in the group of non-deaf users in the First Experiment (with the same trend observable for the rating assigned to the recommendations based on opposite emotions). 

As a consequence of this state of affairs, then, the starting hypothesis guiding this second experiment was confirmed. Indeed, since (in both the experiments) the logic underlying the distribution of the affective-based story recommendations and the actual interface used (i.e. the GAMGame) did not vary, and since the evaluation of recommendations was constrained - by design - to start from stories that that did not represent the original ``first choice" of the users (we remind that they were asked to create brand-new stories considering different items with respect to the one selected in the first experiment, see Procedure)\footnote{Thus representing an even more challenging evaluation setup compared to the first evaluation since the users were, arguably, less incline to provide higher ratings for collections that do not elicit their original preferred emotional setting.}  we arguably attributed the change in the obtained ratings to the effect of a repeated exposition to the overall framework 
 exploring alternative viewpoints vehiculated by stories evoking diverse emotions

This findings is relevant since it suggests that, after a first application of cognitive resistance mechanisms (based on the preservation of one's points of view) the mechanisms of diversity-seeking recommendations seem to be more accepted once the receivers have been already exposed to such type of suggestions. We are not aware of similar results in the context of diversity-seeking recommendations, but these findings are in line with those coming from research on the depolarization of echo-chambers in social media, which show how the repeated exposition to alternative viewpoints, done with techniques like random dynamical nudging, tends to lead the people converge towards less extreme and more inclusive viewpoints \citep{currin2022depolarization}. 

Finally, it is worth observing that, apart from the individual use by the museum visitors, the proposed recommendation system can be used also as a sense-making tool by museum educators and curators in educational activities, where similarities and oppositions can be a relevant tool for guided discussion and comparison in group settings.

\begin{table}

\centering
\begin{tabular}{l|r|rl|} 
\hline
\multicolumn{1}{r}{\textit{Time}} & \multicolumn{3}{c}{\textbf{\textbf{$t_0$}}}                                                                                                                                                                         \\ 
\hline
                                  & \multicolumn{1}{l|}{\begin{tabular}[c]{@{}l@{}}\textbf{Rating similar}\\\textbf{emotion}\end{tabular}} & \multicolumn{1}{l}{\begin{tabular}[c]{@{}l@{}}\textbf{Rating opposite}\\\textbf{emotion}\end{tabular}} &   \\
Mean                              & 6,664                                                                                                  & 5,448                                                                                                  &   \\
Dev.Standard                      & 2,441                                                                                                  & 2,694                                                                                                  &   \\
Median                            & 7                                                                                                      & 5                                                                                                      &   \\ 
\hline
                                  & \multicolumn{3}{c|}{\textbf{\textbf{\textbf{\textbf{\textbf{\textbf{\textbf{\textbf{$t_1$}}}}}}}}}                                                                                                                  \\ 
\cline{2-4}
Mean                              & 7,278                                                                                                  & 7,056                                                                                                  &   \\
Dev.Standard                      & 1,965                                                                                                  & 1,731                                                                                                  &   \\
Median                            & 7                                                                                                      & 7                                                                                                      &   \\
                                  & \multicolumn{1}{l|}{}                                                                                  & \multicolumn{1}{l}{}                                                                                   &   \\
\hline
\end{tabular}
\caption{Rating from the Deaf participants in Experiment 2 at time $t_0$ and $t_1$}
\label{TableDynamicalNudging_ForDeaf}
\end{table}


\subsection{Limitations}
\label{limit}

The overall experimentation faces some limitations. First: as mentioned, the sample of the users that we managed to involve does not allow us to make statistical inferences but only a qualitative analysis on the outcome of the obtained results. Despite this limitation, however, largely due to the difficulty of interacting with such community without the direct intervention of professional experts and translators in Sign language, the number of deaf people involved in the experiment is larger than the typical user studies involving people affected by such a disability. For example, \citeauthor{mack2020social} recruited $7$ participants for interviews on the use of social web apps, and were able to reach a larger sample ($60$ participants) only online. In \citep{mahajan2022towards}, $5$ participants were interviewed.


Second: the experiments reported  in Section \ref{s:evaluation} have been done, differently from the ones reported in \citep{lieto2022degari} on single-items recommendations, outside the museum. As reported, the data of the overall ratings of the first experiment do not seem to suffer from this difference, this suggesting that our system can be used both during a museum visit or in a pre/post visit condition (e.g. for schools).
%
The evaluation described in Section 3, instead, took place in the premises of the museum, but after the visit, since it required a fixed apparatus. However, it is necessary to notice that the GAMGame has not been designed to be employed exclusively during the visit. On the contrary, its design, with its strong visual content and simple, straightforward activities conceived of for inclusion, is suitable for anywhere use, thus contributing to break the barriers between the museum and the outside world, at reach of a  larger number of communities. 


\section{Conclusions and Future Work}

In this paper, we described the use of a novel sensemaking tool for enhancing reflection on cultural items through emotional diversity. 
The system was integrated and evaluated within the context of a Citizen Curation environment.
Targeted to the inclusion of the community of the d/Deaf, this environment allows the user to express their personal interpretations of artworks by creating and sharing simple stories from a collection of museum artworks. 
By leveraging the annotations added by the users to the artworks in the stories, the system allows exploring the repository of stories according to the relations of affective similarity and oppositions between them, as a way to promote perspective taking and empathy.
The results of the evaluation, where the response of deaf users were compared with the non-deaf users, revealed a preference for the stories that are emotionally similar, rather than opposed, to the initial story. For deaf users, this preference tends to emerge with the use, suggesting that the acceptance of this sense-making tool, which contradicts the standard expectations of users, can be improved with the repeated use of the system. 

Despite the encouraging results, as acknowledged in the section \ref{limit}, more data are needed to assess effects of the repeated use of the system on a larger scale. In particular, in future work, we intend to test our approach in the wild, within the context of the activities carried out in the museum by educators and curators with the groups at risk of exclusion, to evaluate its effectiveness as practical tool for involving and including communities, as clearly stated in the 2022 definition of museums released by ICOM.

In particular, it could represent an added value to include in our analysis also deaf children (that were not the target group under investigation in the current project). Extending our analysis in this direction would require a novel testing of the app in order to assess if (and, eventually, to what extent) it is suitable also for this group of interest. In principle, however, the amount of change eventually required should be easily manageable since, as mentioned, the resort to short texts and the adoption of emojis and visual feedback to associate emotional features to museum items should represent a common ground that facilitate the transfer of the testing also to this community.

\section*{Declaration of Competing Interests}
The authors declare that they have no known competing financial interests or personal relationships that could have appeared to influence the work reported in this paper

\section*{Acknowledgements}
The research leading this publication has been partially funded by the European Union’s Horizon 2020 research and innovation programme \url{http://dx.doi.org/10.13039/501100007601} under grant agreement SPICE 870811. The publication reflects the author’s views. The Research Executive Agency (REA) is not liable for any use that may be made of the information contained therein. We thank the GAM Museum and the Istituto dei Sordi di Torino for their help in setting up the evaluation.

\bibliographystyle{apacite}
\bibliography{interactapasample}

\end{document}